\documentclass[showpacs,showkeys, twocolumn,floatfix, prb, aps,superscriptaddress]{revtex4-1}
\usepackage[USenglish]{babel}
\usepackage{fontenc}
\usepackage{graphicx}
\usepackage{amsmath}
\usepackage{epstopdf}
\usepackage{amssymb}
\usepackage{bbm}
\usepackage{amsbsy}
\usepackage{amscd}
\usepackage{float}
\usepackage{color}
\usepackage{standalone}\usepackage{bm}
\usepackage{siunitx}
\usepackage[normalem]{ulem} 
\usepackage{verbatim}
\usepackage{url}
\usepackage{dsfont} 
\usepackage[]{physics}
\usepackage[verbose,hypertexnames=false,bookmarksopenlevel=1,filecolor=blue,
linkcolor=blue,citecolor=blue,urlcolor=blue,pdfstartview=FitH,bookmarksopen,bookmarksnumbered,
colorlinks,plainpages=false,linktocpage]{hyperref}
\DeclareSymbolFont{mathbold}{OML}{cmm}{b}{it}

\newcommand{\id}{\mathbbm{1}} 
\newcommand{\I}{{i\mkern1mu}} 

\newcommand{\hop}{\xi}
\newcommand{\CH}{{\cal H}}
\newcommand{\ka}{\bar{k}}
\newcommand{\betaT}{\bar{\beta}}
\newcommand{\alphaT}{\bar{\alpha}}
\begin{document}
\title{Topological Transitions in two-dimensional Floquet superconductors}
\author{Paul Wenk}
\email{wenkp@wwu.de}
\affiliation{Institute for Physical Chemistry, University of Münster, Münster, Germany}
\author{Milena Grifoni}
\author{John Schliemann}
\affiliation{Institute for Theoretical Physics,
  University of Regensburg, Regensburg, Germany}
\date{\today}
\begin{abstract}
  We demonstrate the occurrence of a topological phase transition
  induced by an effective magnetic field in a two-dimensional electron
  gas with spin-orbit coupling and in proximity to an $s$-wave
  superconductor. The effective, perpendicular magnetic field is
  generated by an in plane, off-resonant ac-magnetic field or by
  circularly polarized light. The conditions for entering the
  topological phase do not rely on fine parameter tuning: For fixed
  frequency, one requires a minimal amplitude of the effective field
  which can be evaluated analytically. In this phase, chiral edge
  states generally emerge for a system in stripe geometry unless the
  Rashba and Dresselhaus coupling have the same magnitude. In this
  special case, for magnetic field driving the edge states become
  Majorana flat bands, due to the presence of a chiral symmetry; the
  light irradiated system is a trivial superconductor.
\end{abstract}
\maketitle
\section{Introduction}%
\label{sec:introduction}
%
Nowadays pathways to manipulate material properties by a time-periodic
drive are often dubbed as ``Floquet engineering''~\cite{Oka_2019}.
This is based on the observation that the time evolution and steady
state of a quantum system under time-periodic driving can be described
in terms of a Floquet Hamiltonian, whose quasi-eigenenergy spectrum
can be entirely different from the spectrum of the undriven
Hamiltonian~\cite{Shirley1965, Grifoni1998}. The growing interest in
this field is fueled by rapid developments in optical and microwave
coherent control techniques, together with the possibility of
engineering novel quantum materials exhibiting exotic electronic
properties. For example time- and angle-resolved photoemission spectroscopy
\cite{Bovensiepen_2012} has been used to image the
Floquet-Bloch surface states of a topological insulator, and to
demonstrate that the surface Dirac cone becomes gapped upon
irradiation by circularly polarized
light~\cite{10.1126/science.1239834}. Vice versa, intense circularly
polarized light might turn a trivial static conductor like graphene
into a Haldane's Chern insulator supporting chiral edge
modes~\cite{Kitagawa2011}.

\begin{figure}[t]
	\includegraphics[width=1\linewidth]{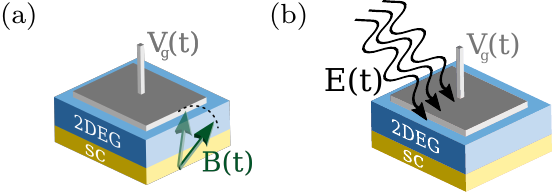}
	\caption{A 2DEG in proximity to a conventional superconductor
          can become itself a (synthetic) superconductor. The combined
          action of spin-orbit coupling (SOC) and a time-dependent
          magnetic (a) or electric (b) field can drive the system to a
          topologically non trivial phase.}%
	\label{fig:sketch}
\end{figure}

While Floquet topological insulators are by now largely understood,
the study of Floquet topological superconductors, in the focus of this
work, is still in an early stage.  \textit{Static} topological
superconductors have attracted a great deal of interest for the
realization of Majorana fermions in solid states and their possible
application to topological quantum
computation~\cite{Sato_2017}. Crucially, the target system should have
spin nondegenerate bands, e.g.\ due to the combined effects of
spin-orbit coupling (SOC) and of a magnetic field. The possibility of
tuning topological superconductivity by light has been discussed by
now in various one-dimensional setups ranging from cold atom chains to
Rashba nanowires~\cite{Jiang_2011, Reynoso2013, Thakurathi2013,
  Liu_2013, Kundu2013, Thakurathi_2017}. Some studies have addressed
the possibility to observe Floquet-Majorana modes in finite
two-dimensional (2D)
systems, like the Kitaev model on honeycomb
lattices~\cite{Thakurathi2014}, a two-band model with $s$-wave pairing
on a square lattice~\cite{Poudel2015}, or a square model with $d$-wave
pairing for cuprate superconductivity~\cite{Takasan_2017}. By applying
off-resonant drive modulating some of the system's properties
(chemical potential, spin-orbit coupling strength, etc.), edge modes
have been predicted based on numerical diagonalization of the
associated Floquet Hamiltonians.  In a recent work,
Plekhanov~\textit{et al.}~\cite{Plekhanov_2019} have posed the
question, whether a Floquet topological phase transition can be
entered also in the more conventional set-up of a two-dimensional
electron gas (2DEG) with spin-orbit coupling being proximity coupled
to an $s$-wave superconductor. Under resonance conditions between two
spin-orbit split bands, an out-of plane magnetic field, see
Fig.~\ref{fig:sketch}(a), is predicted to induce helical edge
modes. Despite appealing, a limitation of this proposal is that the
resonant condition can be satisfied only in a restricted region in
$k$-space. Where the resonance is not satisfied, more and more Floquet
subbands participate in the low energy behavior and the topological
gap disappears.

Motivated by the former studies on 2D Floquet
superconductors~\cite{Poudel2015,Takasan_2017,Plekhanov_2019}, we
investigate here the impact of off-resonant time-dependent
electromagnetic fields for the setups in Fig.~\ref{fig:sketch}.
Although we prevalently focus on the one in Fig.~\ref{fig:sketch}(a),
the two-models map onto similar effective Hamiltonians obtained by
using L\"owdin partitioning. We find that a high-frequency in plane
ac-magnetic field generates an out-of plane component of an effective
static magnetic field which can drive the topological phase transition. Our analytical results
are corroborated by diagonalization of the full Floquet Hamiltonian
for bulk as well as stripe geometries. The topological phase
boundaries can be calculated analytically, and are well in agreement
with numerical predictions for topological invariants of the bulk
system. In turn, the Floquet spectrum of a finite stripe displays
chiral edge modes crossing at the $\Gamma$ point, for generic values
of the SOC, in the parameter region of finite Chern
number. Remarkably, flat bands are found in the limit of equal
amplitude but different sign of Rahsba and Dresselhaus SOC.\@ They are
enforced by a chiral symmetry of the driven system~\cite{Deng2014}
present in this special case.  For the laser irradiated setup, the
effective out-of-plane magnetic field vanishes when the Rashba and
Dresselhaus SOC have the same amplitude. In this case there is no
topological phase. Finally, an interesting scaling of the topological phase transition with
system size is observed.

The paper is structured as follows: In Sec.~\ref{sec:model} we construct the
Floquet-Bogoliubov de Gennes (BdG) Hamiltonian matrix for our model,
while in Sec.~\ref{sec:high-freq-limit} the high frequency effective Hamiltonian is derived
within L\"owdin perturbation theory~\cite{BirPikusBook,winklerbook}. In Sec.~\ref{sec:edge-modes-stripe} numerical results are
presented for stripe geometries and compared with the expectation of
the effective static model. Finally, conclusions are drawn in
Sec.~\ref{sec:conclusions}. Details of the calculations are presented
in Appendices \ref{sec:floq-bdg-hamilt}--~\ref{sec:char-topol-phase}.

\section{Model}%
\label{sec:model}
With focus on the setup of Fig.~\ref{fig:sketch}(a) we consider a 
continuum Hamiltonian near the $\Gamma$ point
\begin{align}
  {\cal H}(t)={\cal H}_0+{\cal H}_\Delta+{\cal H}_1(t)\;,
  \label{general1}
\end{align}
where the static part describes a 2DEG with
Rashba~\cite{rashba_1,Rashba2015SymmetryOE} ($\alpha$) and linear
Dresselhaus~\cite{Dresselhaus1955} ($\beta$) SOC in proximity to an
$s$-wave superconductor. Explicitly,
\begin{align}
{\cal H}_0 = & \nonumber\\  
\sum_{\sigma\sigma^{\prime}} \int & d^2k \, 
  \psi^\dagger_{\vb{k} \sigma}
  {\left(\frac{\hbar^2k^2}{2m^*}-\mu
  -\alpha_x k_x\sigma_y+\alpha_y k_y\sigma_x\right)}_{\sigma\sigma^{\prime}}
  \psi_{\vb{k}\sigma^{\prime}},\nonumber\\
  {\cal H}_\Delta = & -\frac{\Delta}{2}\sum_{\sigma\sigma^{\prime}}\int d^2k\, 
  \left(\psi^\dagger_{\vb{k} \sigma}{(i\sigma_y)}_{\sigma\sigma^{\prime}}
  \psi^\dagger_{-\vb{k}\sigma^{\prime}}+{\rm h.c.}\right)\,.
  \label{general2}
\end{align}
Here, $\psi^\dagger_{\vb{k} \sigma} (\psi_{\vb{k} \sigma})$ are
creation (annihilation) operators of an electron with spin component
$\sigma$ along the $z$ direction and wave vector $\vb{k}$; for the SOC
constants we define $\alpha_x:=\alpha + \beta$,
$\alpha_y := \alpha - \beta$, and in real as well as spin space the
$x$-coordinate axis points along the crystallographic
$[1,\bar 1,0]$-direction while the $y$-axis lies along $[1,1,0]$;
$m^*$ is the effective mass and $\mu$ the chemical
potential. Superconductivity induced in the 2DEG by proximity effects
is captured by the mean-field Hamiltonian ${\cal H}_\Delta$, where
$\Delta$ is the proximity induced superconducting gap. For the
time-dependent part driving the spin dynamics in the 2DEG we consider
the generic expression
\begin{align}
 {\cal H}_1(t) =
\sum_{\sigma\sigma^{\prime}}\int d^2k\,
  \psi^\dagger_{\vb{k} \sigma}
  {\left(\vb{A}(t)\cdot \vb*{\sigma}
  	\right)}_{\sigma\sigma^{\prime}}
  \psi_{\vb{k}\sigma^{\prime}}\,.
  \label{general3}
\end{align}
Here, $A_i(t)=\mu_B (g_i/2)B_i(t)$, with $g_i$ the effective
gyromagnetic ratio along the $i$-direction, the magnetic field
amplitude
\begin{align}
  B_i(t)=B_{ic}\cos (\Omega t)+B_{is}\sin(\Omega t)\;,
\end{align}
$\vb*{\sigma}=(\sigma_x,\sigma_y,\sigma_z)$ the vector of Pauli
matrices. Thus, in full generality,
\begin{align}
  \vb{A}(t) = \vb{q}\cos(\Omega t) + \vb{r}\sin(\Omega t)\qq{with} \vb{q}, \vb{r} \in \mathbb{R}^3.
\end{align}
%
%

Due to the periodicity of the external magnetic field, we use Floquet
theory to find the quasi-energy spectrum and discuss topological
properties. Explicitly, the time dependent problem can be recast onto
an effective static problem for the Floquet Hamiltonian
\begin{align}
  {\cal H}_F(t)= {\cal H}(t)-i\hbar \partial_t
\end{align}
when working in the
composite Sambe space $\cal{S}=\mathbb{T} \otimes \mathbb{H}$ spanned
by time periodic functions and the conventional Hilbert space
${\mathbb H}$~\cite{Oka_2019,Grifoni1998}. We introduce the matrix
elements in ${\mathbb T }$
of ${\cal H}(t) $
 in terms of the Fourier transform 
\begin{align}
{\cal H}_{mn} = (m\vert {\cal H}(t)\vert n):= & \frac{1}{T} \int_0^T d t\, {\cal H}(t) e^{\I (m-n)\Omega t}\, .
\label{eq:Floquet}
\end{align}
In turn the Floquet Hamiltonian has matrix elements
\begin{align}
  {({\cal H}_F)}_{mn} := {\cal H}_{mn} -
n\hbar\Omega\delta_{mn}\;.
\end{align}
Diagonalization of ${\cal H}_F$ then yields
the quasienergy spectrum of the driven system.  Care has to be taken
when including the contribution of the mean field term
${\cal H}_\Delta$ in Eq.~\eqref{eq:Floquet}. Due to the presence of
two creation or two annihilation operators, the superconducting term
is off-diagonal in Sambe space and couples Fourier modes $n$ and -$n$.

Similar to the case of static superconductors, the evaluation of the
quasi-energy spectrum is more conveniently performed introducing the
Nambu spinor
$\Psi^\dagger_{\vb{k}} =
(\psi^\dagger_{\vb{k}\uparrow},\psi^\dagger_{\vb{k}\downarrow},
\psi_{-\vb{k}\uparrow},\psi_{-\vb{k}\downarrow})$. Then ${\cal H}(t)$
can be expressed in terms of BdG matrices,
\begin{align}
  {\cal H}(t) ={}&  \frac{1}{2}\int d^2k\, \Psi^\dagger_{\vb{k}}(H_0(\vb{k})+ H_\Delta + H_1(t))\Psi_{\vb{k}}\;.
  \label{general4}
\end{align}
The three parts of the BdG Hamiltonian are given by~\footnote{Tensor
  convention: $\tau_z\otimes\sigma_i = \begin{pmatrix}
\sigma_i & \vb{0}_2\\
\vb{0}_2 & -\sigma_i
  \end{pmatrix}$}
\begin{align}
  H_0 ={}& (\varepsilon_k-\mu)\tau_z\otimes\id_2 +
           \alpha_y k_y\id_2\otimes\sigma_x -\alpha_x
           k_x\tau_z\otimes\sigma_y ,\nonumber\\
H_\Delta = {} &  \Delta\tau_y\otimes \sigma_y\;, \label{BdGHamilParts}\\
  H_1(t) ={}& A_{z}(t)\tau_z\otimes\sigma_z +
A_x(t)\tau_z\otimes\sigma_x +
A_y(t)\id_2\otimes\sigma_y\;,\nonumber
\end{align}
with $\varepsilon_k =\hbar^2k^2/2m^*  $ and $\tau_i$ Pauli matrices in Nambu space.  
%
Together with Eq.~\eqref{eq:Floquet}, this leads to the Floquet-BdG
matrix
\begin{align}
  {(H_F)}_{mn} :={}& H_{mn} - n\hbar\Omega\delta_{mn}
\tau_z\otimes\id_2\;.
\end{align}
Diagonalization of $H_F$ then yields the
quasienergy spectrum of the driven system. Its form is
provided in Appendix~\ref{sec:floq-bdg-hamilt}. While for
generic driving frequencies a numerical diagonalization is required,
analytical results can be obtained for off-resonant driving, the case
of interest in this work.
\section{High-Frequency Limit: L\"{o}wdin Partitioning}%
\label{sec:high-freq-limit}
High-frequency driving is known to be an excellent tool to dress
parameters of the static, unperturbed Hamiltonian, e.g.\ hopping
elements or on-site energies~\cite{Sheremet2016}, leading
to phenomena like coherent destruction of tunneling or dynamical
localization~\cite{Grifoni1998}, which can be used to steer a
topological phase transition~\cite{Liu_2013_Floq}.  In addition,
off-resonant drive can generate terms absent in the static
Hamiltonian~\cite{Takasan_2017}, a crucial effect in the
following. Starting from the Floquet matrix ${(H_F)}_{mn}$, we
truncate it by retaining only the $\{m,n= 0,\pm 1\}$ blocks (of
dimension $4\times4$) into account. This gives rise to three Floquet
modes. The truncation is justified when $W/\hbar\Omega$ is a small
quantity, with $W$ being the static bandwidth. We then apply the
L\"{o}wdin perturbation scheme~\cite{BirPikusBook,winklerbook},
allowing us to fold the effects of terms including the higher
harmonics $\pm 1$ in an effective static model with only the $n=0$
block.
Explicitly, see Appendix~\ref{sec:lowd-high-freq}, L\"{o}wdin perturbation  
 up to first order in $W/\hbar\Omega$ results in the 
effective BdG Hamiltonian
\begin{align}
  H_{\text{eff}} ={}& H_0 + H_\Delta + \frac{\Lambda_z}{\hbar\Omega}\tau_z\otimes\sigma_z +
     \frac{\Lambda_y}{\hbar\Omega}\id_2\otimes\sigma_y+
     \frac{\Lambda_x}{\hbar\Omega}\tau_z\otimes\sigma_x 
                      \label{Heff_1}
\end{align}
with $\vb*{\Lambda}=\vb{r}\times \vb{q}$. Notably, the \emph{in plane}
components of the magnetic field generate an \emph{out of plane} term
in the effective Hamiltonian. This feature is crucial for the
occurrence of the topological phase transition. Hence, in the following we choose $B_z=0$,
such that $\Lambda_x=\Lambda_y=0$. Introducing the Zeeman term
$h_{\text{Z}} := \Lambda_z /\hbar\Omega$, the effective BdG
Hamiltonian reduces to the appealing form
\begin{align}
  H_{\text{eff,Z}} ={}& \qty(\varepsilon_k  - \mu )\tau_z\otimes\id_2 
            -\alpha_x 
           k_x\tau_z\otimes\sigma_y\nonumber\\
{}& +\alpha_y 
k_y\id_2\otimes\sigma_x + \Delta\tau_y\otimes \sigma_y + h_{\text{Z}}\tau_z\otimes\sigma_z\;.
\label{Heff_Z}
\end{align}
The spectrum of $H_{\text{eff,Z}}$ can be evaluated in closed form.
Importantly, the superconducting term exhibits a $p$-wave component in
the eigenbasis of the normal conducting ($\Delta=0$) system, see
Appendix~\ref{sec:floq-bdg-hamilt}. The gap closing condition yields the critical Zeeman
amplitude $h_{\text{Z},c}$ which separates the topologically trivial
from the non-trivial phase. For $|\alpha|\neq|\beta| $ it is
\begin{align}
  |h_{\text{Z},c}| = \sqrt{\Delta^2 + \mu^2}\;.\label{h_critical}
\end{align}
This result is remarkably simple, independent of the SOC strength, and
closely resembles the one for some Rashba 1D~\cite{Alicea_2012} and
2D~\cite{Sau_2010} setups. It implies that chiral edge modes should
appear in a finite size system, under appropriate choice of the
system's parameters. 
\section{Edge modes in stripe geometries}%
\label{sec:edge-modes-stripe}
The emergence of such chiral modes is confirmed from a numerical
evaluation of the quasi-energy spectrum of the Floquet-BdG Hamiltonian
$H_{F}$ in stripe geometry. Further, topological invariants for the
bulk system were calculated numerically and also agree with the
expectations of the simple high frequency model. The numerical results
can be seen in
Figs.~\ref{fig:toptrans4}--\ref{fig:flatbandsstability}, and are
commented in detail below.
\subsection{$\vert \alpha \vert \neq \vert \beta \vert$, chiral edge modes}\label{sec:alphaNEQbeta}
For generic values of the spin-orbit coupling parameters $\alpha$ and
$\beta$ chiral modes are expected to emerge in the topological
phase. This is observed in Figs.~\ref{fig:toptrans4}(a) and
~\ref{fig:toptrans4}(b). Further, Fig.~\ref{fig:toptrans4}(c)
demonstrates the exponential decay of the modes towards the interior
of the stripe.  The prediction of a topological boundary given in
Appendix~\ref{subsec:specHeff} by Eq.~\eqref{h_critical} [red line in
Fig.~\ref{fig:toptrans4}(d)] agrees with a numerical evaluation of the
Chern numbers $C_\alpha$~~\cite{Thouless1982a,Berry1984,Simon1983}
associated to the four quasienergy bands $\varepsilon_{\vb{k}\alpha}$,
$\alpha=1,2,3, 4$ of the central Floquet zone of the bulk Floquet Hamiltonian $H_F$ (see Appendix~\ref{app:FloquetTB}). 
As seen in Fig.~\ref{fig:toptrans4}(d), $C_2=0,1$   
before and after
the gap closing at $h_{\text{Z},c}$. 
Because the phase boundary only depends on the ratio $h_Z/\Delta$ and
$\mu/\Delta$, it does not change if these quantities are scaled by the
same factor. For the edge modes to appear though the width of the
system should be appropriately changed, such that the decay lengths of
the modes remains much smaller that the width. This property
can be verified from the figures shown in the Appendix.\@ Here, the relevant
scales are a factor 20 larger; edge modes appear for correspondingly
narrower stripes.
\subsection{$\vert \alpha \vert = \vert \beta \vert$, Majorana flat bands}\label{sec:majorana-flat-bands}
The above results are quite generic and hold true for off-resonant
driving and $|\alpha|\neq|\beta| $. Hence they do no require fine
tuning of
parameters. 
In the following we discuss one of the special symmetry setting
supporting persistent spin
helices~\cite{Schliemann2003,Koralek2009,Kammermeier2016a,Schliemann_2017},
$|\alpha|=|\beta| $. Here, a special choice of parameters leads to
Majorana flat bands~\cite{Oshima_2022}. This can occur in two
situations. In case I the Rashba and Dresselhaus SOC strength have
opposite sign $\alpha=-\beta$, which implies $\alpha_x=0$, and
$\alpha_y=2\alpha$. Then, Majorana flat bands are found in stripe
geometry with finite width along the $y$-direction. In the case II,
$\alpha=\beta$, the flat bands occur for a finite width of the stripe
in the $x$-direction. We demonstrate below that the emergence of flat
bands is strictly related to the presence of an additional chiral
symmetry of the driven system.

A magnetic field breaks time-reversal symmetry. Hence, for a generic
parameter set, also the effective BdG Hamiltonian Eq.~\eqref{Heff_Z}
is only invariant under particle-hole symmetry, with
${\cal P}=\tau_x\otimes\id_2{\cal K}$ the associated antiunitary
particle-hole operator and ${\cal K}$ the operator of complex
conjugation; it holds ${\cal P}^2=1$. In this case the system belongs
in 2D to the symmetry class D~\cite{Altland_1997}, and its topological
properties are well described in terms of Chern numbers, as discussed
above. For the special case I,
\onecolumngrid{}

\begin{figure}[t]
	\centering
	\includegraphics[width=\linewidth]{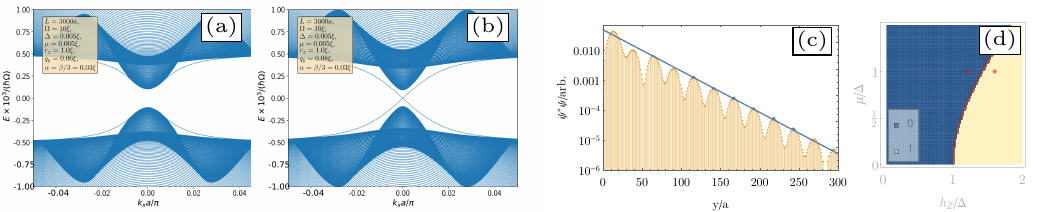}
	\caption{Topological phase transition induced by an in plane time-dependent magnetic
          field. (a)-(b) Numerical quasienergy spectrum of $H_{F}$ in
          stripe geometry for the two values of chemical potential and
          Zeeman field shown with red dots in panel (d). We considered
          $N_y=3600$ transverse channels and retained $N_F=3$ Floquet
          modes. In the topological phase, panel (b), chiral mode
          crossing at the $\Gamma$ point emerge. (c) The two chiral
          edge modes decay exponentially towards the device interior
          and are localized at opposite edges. Here one of them is
          shown for $k_x a=0.1$.  (d) Chern number $C_2$ of the   bulk
          Floquet-BdG Hamiltonian   for different values of  $h_{Z}$
          and
          $\mu$. It holds $C_2=1$  in the nontrivial region; its boundary, depicted by the red solid line, is given by Eq.~\eqref{h_critical}. 
        }%
	\label{fig:toptrans4}
\end{figure}
\vspace{-0.8cm}
\begin{figure}[t] 
	\centering
	\includegraphics[width=\linewidth]{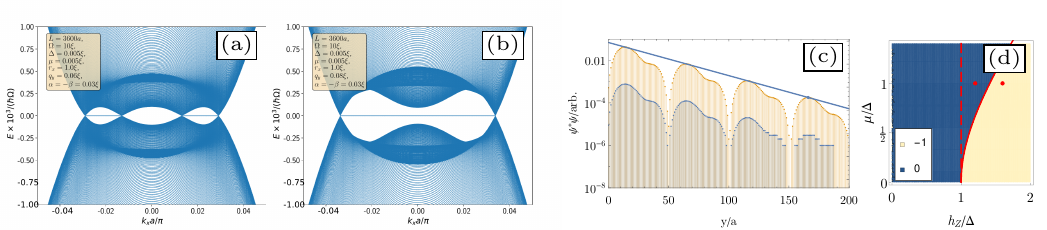}
	\caption{Majorana flat bands for the symmetric case with SOC
		parameters $\alpha=-\beta$. 
		 (a)-(b) Quasienergy spectrum of $H_F$ in stripe
                 geometry with Zeeman amplitude and chemical potential
                 corresponding to the two red dots on the left (a) and
                 right (b) of the topological phase boundary [solid red line  in panel (d)]. 
                 (c) The spatial profile of the Majorana modes is
                 shown for $k_x a = 0.1 $.  (d) The Partial
                 Berry-phase sum parity $P_{B,k_x=0}=P_{B,k_x=\pi/2}$
                 is shown as a function of $h_Z/\Delta$ and
                 $\mu/\Delta$.  
                 The dashed vertical line separates the region with
                 and without flat bands.}\label{fig:toptrans5}
\end{figure}
\twocolumngrid{} 
\noindent
$\alpha=-\beta$, the effective
Hamiltonian acquires a chiral symmetry with
${\cal C}_{\rm I}=\tau_y\otimes \sigma_z$ the associated unitary
operator obeying ${\cal C}_{\rm I}^2=1$. As a consequence, the high
frequency BdG Hamiltonian is also invariant under the time-reversal
operation generated by ${\cal T}_{\rm I}={\cal P}{\cal C}_{\rm
  I}$. Notice that ${\cal T}_{\rm I}^2=-1$.  In this case the system
belongs to the symmetry class DIII;\@ it can support flat bands if the
associated $\mathbb{Z}_2$ invariant becomes nontrivial~\cite{Sato_2017}. Similarly, we find for case II, $\alpha=\beta$, the
chiral symmetry ${\cal C}_{\rm II}=\tau_x\otimes\id_2$.  The gap can
now close at finite momenta. Introducing polar coordinates
$k_x=k\cos\theta$, $k_y=k\sin\theta$, the closing occur when
$\theta=0,\pi/2$ for case I, II respectively, and
 \begin{align}
 |h_{Z,c}(k)| ={}& \sqrt{\Delta^2 + {\qty(\mu-a^2k^2\xi)}^2} \,,
 \label{eq:h_effK}
 \end{align}
where $\xi$ is the hopping energy and $a$ the lattice constant in the square lattice tight-binding formulation. 
In Fig.~\ref{fig:toptrans5} the appearance of flat bands hosting Majorana edge modes   is shown for case I and a stripe with finite width along the $y$-direction. Notice that only the amplitude of the spin-orbit coupling strength was changed compared to Fig.~\ref{fig:toptrans4}. 
Also, no flat bands are found in case II.\@ The situation is reverted
if the stripe has a finite geometry along the $x$-direction,
reflecting the directionality of the effective SOC field.  According
to Eq.~\eqref{eq:h_effK}, see Fig.~\ref{fig:toptrans5}(a), situations
may happen where flat bands only exist in certain regions of
$k$-space. This mixed regime occurs in the parameter region between
the topological phase boundary and the vertical red line in
Fig.~\ref{fig:toptrans5}(d). We expect the mini-flat bands not to be
stable against perturbations.  Here, as ${\mathbb{Z}_2}$ invariant the partial Berry phase numbers proposed in Ref.~[\onlinecite{Deng2014}] were used, see Sec.~\ref{sec:partial-berry-phase} of the Appendix.\\
\subsection{Topological signatures beyond off-resonance}%
\label{topres}
The topological phase transition and the associated edge modes in
finite geometries have been obtained for driving energies
$\hbar\Omega$ much larger than the bandwidth $W$ of the static 2D
system.  
In Fig.~\ref{fig:flatbandsstability} we show that topological features
are clearly discerned also for $\hbar \Omega =7\ \hop$, where $\hop$
is the hopping energy in the tight-binding formulation. A series of
crossing and avoided crossing is observed not only for the in gap
modes, but also for the higher energy. Such features are common to
other topological systems and are a signature of the topological
character of the excitations near the avoided crossing, see,
e.g., 
Refs.~[\onlinecite{Mishmash2016,Leumer_2021}].
\section{Experimental feasibility and conclusions}%
\begin{figure}[t]
	\centering
	\includegraphics[width=\linewidth]{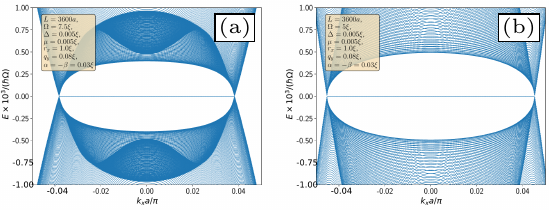} 
	\caption{Stability of flat bands. Midgap states and Majorana
          oscillations are still seen by lowering the driving
          frequency from $\hbar\Omega=7.5~\hop$ in panel  (a) to $\hbar\Omega=5~\hop$  in panel (b). 
        }\label{fig:flatbandsstability}
\end{figure}

\label{sec:conclusions}
We have demonstrated the emergence of topologically non-trivial phases
in a theoretical model of a 2DEG subject to an in plane off-resonant
magnetic field. A natural question is to which extent such phases can
be observed in the state of the art experimental set-ups. For a 2DEG from
ordinary III-V semiconductors we estimate a hopping $\hop\simeq 1.5$~eV, which
implies frequencies in the energy range $\hbar \Omega = 5-15$~eV for
the plots in Figs.~\ref{fig:toptrans4}
---~\ref{fig:flatbandsstability}. Further, with
$\Delta \simeq$~0.1~meV and a gyromagnetic ratio $g =50$ (e.g.\ for
InSb~\cite{Qu_2016}), one finds magnetic field amplitudes
$B_{xs}\simeq B_{yc} \approx 30$~T. These are rather large and destroy
superconductivity, if the latter is induced through proximity to a
conventional superconductor like Nb having critical fields in the
order of 0.8~T.  A possible extension of this work thus points to
proximity to 2D Ising superconductors, like e.g. NbSe$_2$, which are
known to support large in plane critical fields of more than 30~T
\cite{Xi_2016}. 
Signatures of triplet superconductivity were recently observed in
trilayer NbSe$_2$ driven by a static in plane magnetic field up to
33~T~\cite{Kuzmanovic_2021}. The pairing function discussed there has
similar $s$- and $p$-wave components as derived in Appendix \ref{sec:floq-bdg-hamilt}.\@ This suggests related
low-energy physics for seemingly distinct 2D superconductors.
Alternating magnetic fields with amplitudes of several Tesla are also
difficult to achieve in ordinary laboratories. Hence, we would like to
comment also on the setup in
Fig.~\ref{fig:sketch}(b), with the 2DEG driven by circularly polarized
light with vector potential
${\cal A}(t)=({\cal A}\cos (\Omega t),{\cal A}\sin(\Omega t),0)$.  In
this case, the light couples to the 2DEG electrons through a minimal
coupling ${\bf{k}}\to {\bf{k}}-{\bf{k}}_0(t)$, with
${\bf{k}}_0(t)=(e/c\hbar){\cal A}(t)$. Following Refs.~[\onlinecite{Mikami2016,Takasan_2017}], in the off-resonant case, L\"owdin
perturbation theory leads again to the Hamiltonian Eq.~\eqref{Heff_Z},
with the replacement $m^*\to m^*/J_0(x)$,
$\alpha_{x,y}, \to \alpha_{x,y} J_0(x)$ in the single particle part
and $\Lambda_z\to - (\alpha_x\alpha_y/a^2) J_1^2(x)$. Here, $J_n(x)$
is a Bessel function of first kind and $x=(ea/c\hbar) {\cal
  A}$. Driving by light has the advantage that, since
$\vert J_0\vert \leq 1$, the hopping $\hop \simeq 1/m^* $ gets
effectively reduced and hence also the frequencies $\Omega $ being
required.
Also for this model chiral modes are expected for effective magnetic
fields $h_{Z}$ larger than the critical field Eq.~\eqref{h_critical}
and generic SOC parameters.  However, in the symmetric cases
$\alpha=\pm\beta$ is either $\alpha_x=0$ or $\alpha_y=0$, leading to a
vanishing $h_Z$ and thus to trivial superconductivity.\\
In summary, in our work we focused on proximitized semiconducting
2DEGs with strong spin-orbit coupling as possible systems for the
realization of a Floquet topological superconductor. However, the
large amplitudes of the magnetic/electric fields and the high
frequencies necessary to induce Floquet topological superconductivity
indicate the need to look for alternative 2D materials.  We suggest
that 2D superconductors of the Ising type, like few layers NbSe$_2$,
are an ideal candidate to observe the topological phase transition. Since
off-resonant magnetic fields induce effective static magnetic fields,
our analysis can be applied also to situations in which only
appropriate static fields are applied.
\begin{acknowledgments}
This work was supported by the German Science
Foundation under CRC 1277, Project No. B09.  We thank J.\ Klinovaja and
D.\ Loss for discussions.
\end{acknowledgments}
%
\appendix
%
\section{Floquet BdG Hamiltonian in the continuum model}%
\label{sec:floq-bdg-hamilt}
\subsection{BdG Hamiltonian in Sambe space}\label{sec:bdg-hamilt-sambe}
\begin{figure}[t]
  \centering
  \includegraphics[width=\linewidth]{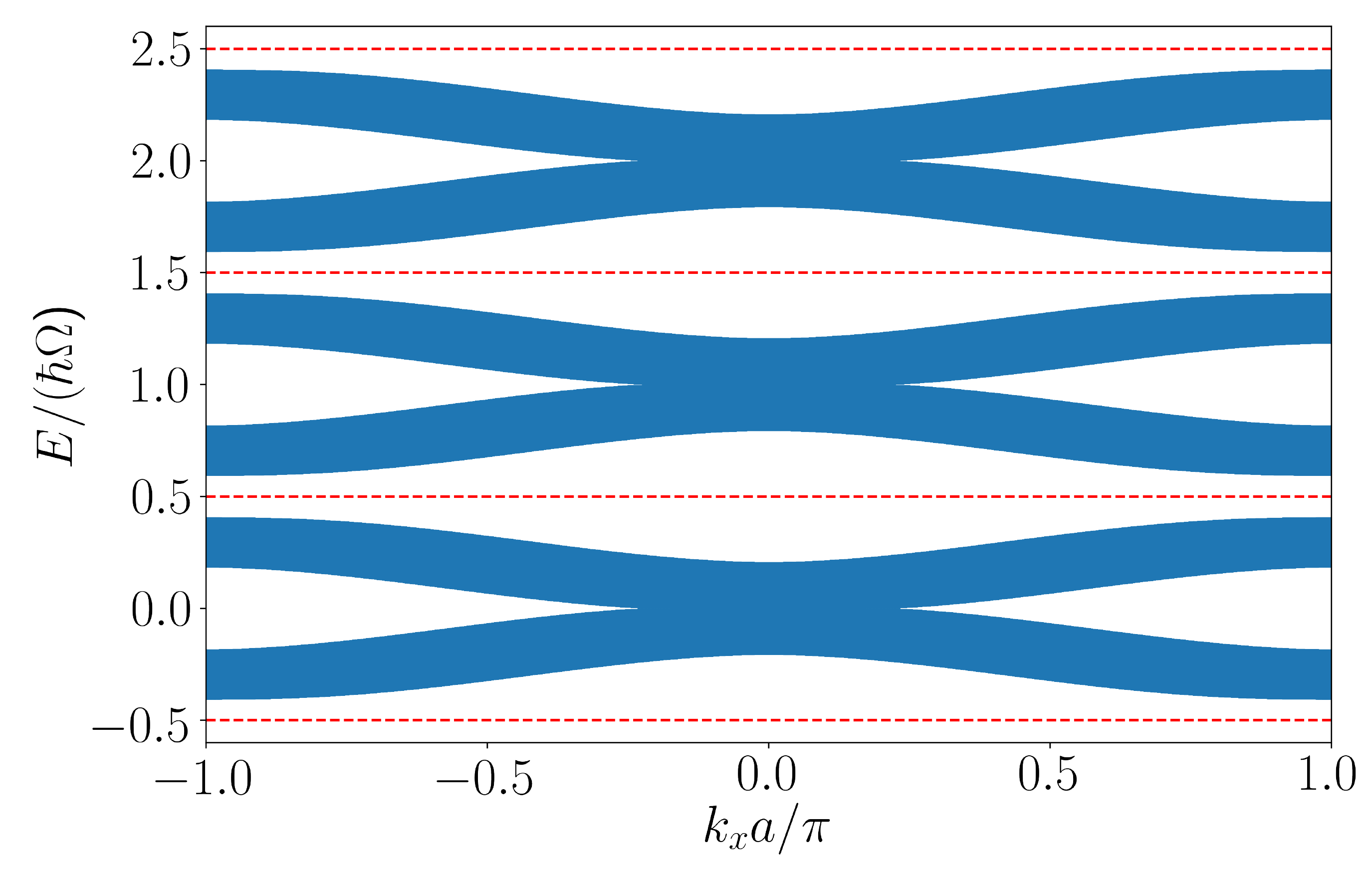}
  \caption{As expected from a Floquet Hamiltonian, replicas appear
    which are separated by the horizontal red dashed lines.}%
  \label{fig:replica}
\end{figure}

We explicitly derive the Floquet-BdG matrix associated to
the continuum model Hamiltonian
\begin{align}
  \CH(t) ={}&  \CH_0 + \CH_\Delta + \CH_1(t)\;,
  \label{general1App}
\end{align}
defined in Eq.~\eqref{general1}. Since $\CH(t)$ is time
periodic, $\CH(t) = \CH(t + T)$, with the driving period
$T=2\pi/\Omega$, we can apply Floquet theory~\cite{Floquet_1883,Grifoni1998}. The solution to the
Schr\"odinger equation
\begin{align}
  i\hbar \partial_t \ket{\phi_\alpha(t)} ={}&  \CH(t)\ket{\phi_\alpha(t)}
\end{align}
are Floquet states
\begin{align}
  \ket{\phi_\alpha(t)} ={}& e^{\frac{\I}{\hbar}\epsilon_\alpha t} \ket{u_\alpha(t)}\;,
\end{align}
with the quasienergies $\epsilon_\alpha$ and the Floquet functions
$\ket{u_\alpha(t)}$. The latter are eigenstates of the Floquet Hamiltonian
$\CH_F(t) := \CH(t) -\I\hbar \partial_t$,
\begin{align}
  \CH_F(t)\ket{u_\alpha(t)} ={}& \epsilon_\alpha \ket{u_\alpha(t)}\;.\label{FloquetEigenvalProb}
\end{align}
Since the Floquet functions have the property of being periodic in
$T$, it is convenient to apply a Fourier expansion
\begin{align}
  \ket{u_\alpha(t)} ={}& \sum_{n=-\infty}^{\infty}
                         \ket{u_\alpha^n}e^{-\I n\Omega t}\;.
\end{align}
Doing the same for $\CH(t)$,
\begin{align}
  \CH_n ={}& \frac{1}{T} \int_{0}^{T}dt\, \CH(t)e^{\I n\Omega t},  
\end{align}
allows one to rewrite the eigenvalue problem in Eq.~\eqref{FloquetEigenvalProb} as
\begin{align}
\sum_{m=-\infty}^{\infty}\underbrace{(\CH_{n-m} -
  n\hbar\Omega\delta_{nm})}_{=:{(\CH_F)}_{nm}} \ket{u_\alpha^m} ={}& \epsilon_\alpha\ket{u_\alpha^m}\;.
\end{align}
Going to Nambu space with
\begin{align}
  \Psi^\dagger_{\vb{k}} ={}& 
(\psi^\dagger_{\vb{k}\uparrow},\psi^\dagger_{\vb{k}\downarrow},
\psi_{-\vb{k}\uparrow},\psi_{-\vb{k}\downarrow})\;,
\end{align}
the Hamiltonian
Eq.~\eqref{general1App} assumes the form 
\begin{align}
  \CH(t) ={}& \frac{1}{2} \int d^2k\, \Psi_{\vb{k}}^\dagger
              {(H_0(\vb{k}) + H_\Delta + H_1(t))}\Psi_{\vb{k}}\;,
\end{align}
where $\CH_{\text{BdG}}(t) := H_0(\vb{k}) + H_\Delta + H_1(t)$ is a
periodic BdG Hamiltonian.
The static part, see Eq.~\eqref{BdGHamilParts},
is given by 
\begin{align}
  H_0 + H_\Delta ={}& \qty(\frac{\hbar^2k^2}{2m^*}  - \mu )\tau_z\otimes\id_2 +
           \alpha_y k_y\id_2\otimes\sigma_x\nonumber\\ 
{}& -\alpha_x k_x\tau_z\otimes\sigma_y
           + \Delta\tau_y\otimes \sigma_y\;, 
\end{align}
and the driving term of general (co-) sinusoidal behavior writes as
\begin{align}
   H_1(t) ={}& \begin{pmatrix}
    h_1(t) & \vb{0}_{2\times 2}\\
    \vb{0}_{2\times 2} & -h_1^*(t)
  \end{pmatrix}\label{driving}\;.
\end{align}
We consider the general form $h_1(t) = \vb{A}(t)\cdot \vb*{\sigma}$
for the time dependent driving, with
\begin{align}
\vb{A}(t) ={}& \vb{q}\cos(\Omega t) + \vb{r}\sin(\Omega t)\\
={}& 
               \qty(\frac{\vb{q} - \I \vb{r}}{2})e^{\I\Omega t} + \qty(\frac{\vb{q} + \I \vb{r}}{2})e^{-\I\Omega t}\;, 
  \label{periodic1c}
\end{align}
which is periodic in time with a period $T=2\pi/\Omega$ and has
parameters $r_l,q_l\in \mathbb{R}$, $l\in\{x,y,z\}$. The crucial part
lies now in writing down the BdG Hamiltonian in Sambe space. Here, one should keep
in mind that annihilation and creation operators are adjoints of each other
and should be Fourier expanded consistently. Thus,
\begin{align}
  \psi_{\vb{k}\sigma}(t) = \sum_{n=-\infty}^{\infty}e^{-\I n\Omega t}\psi_{\vb{k}\sigma,n}\, ,\,\,
  \psi_{\vb{k}\sigma}^\dagger(t) = \sum_{n=-\infty}^{\infty}e^{\I n\Omega
  t}\psi_{\vb{k}\sigma,n}^\dagger\,,
  \label{periodic2}
\end{align}
in the Nambu spinor. Including such time dependence has the consequence that the superconducting term parametrized by
$\Delta$ couples Fourier modes $n$ and $-n$ in the Floquet equation to
be discussed below. The correct anti-diagonal position of the
superconducting gap is crucial for the quasienergy spectrum to exhibit
the usual appearance of replica, which we will explicitly show for the
stripe geometry, Fig.~\ref{fig:replica}. 
To be more precise, in the Nambu-Sambe basis $(\ldots,\Psi_{\vb{k},-N}^\dagger,
\Psi_{\vb{k},-N+1}^\dagger,\ldots,\Psi_{\vb{k},N}^\dagger,\ldots)$ we
find the Floquet-BdG matrix
\begin{align}
  {(H_F)}_{mn} ={}& H_{mn} -n\delta_{mn}\hbar\Omega\tau_z\otimes\id_2\;.
\end{align}
In the following $H_F$ is written down in a truncated form with $N$ Floquet
modes,
\begin{widetext}
  \begin{align}
    H_F ={}& \begin{pmatrix}
               H_0 + N\hbar\Omega\, \tau_z\otimes \id_2 & d                                 &
                                                                                              \vb{0}_{4\times4} & \cdots & \vb{0}_{4\times4}\\
               d^\dagger                        & H_0 +
                                                  (N-1)\hbar\Omega\, \tau_z\otimes \id_2 & d & \cdots & \vb{0}_{4\times4}\\
                                                        &    \ddots         & \ddots & \ddots &\\
               \vb{0}_{4\times4} & \cdots  & d^\dagger                        & H_0 -
                                                                                (N - 1)\hbar\Omega\, \tau_z\otimes \id_2 & d\\
               \vb{0}_{4\times4} & \cdots & \vb{0}_{4\times4} & d^\dagger & H_0 -
                                                                            N\hbar\Omega\, \tau_z\otimes \id_2
             \end{pmatrix}\nonumber\\
    {}& + \Delta\begin{pmatrix}
                  & & & s \\
                  & & \reflectbox{$\ddots$} & \\
                  & s & & \\
                  s & & &  \\
                \end{pmatrix}\;,\label{FullFloquetH}
  \end{align}
\end{widetext}
with the matrices
\begin{align}
  d ={}& \begin{pmatrix}
h_1^1 & 0\\
0 & -{(h_1^1)}^*
  \end{pmatrix}\;,\label{d_matrix}
\end{align}
\begingroup
\allowdisplaybreaks{}
\begin{align}
h_1^1 ={}& \frac{1}{2} (\vb{q} - \I \vb{r})\cdot \vb*{\sigma}\\ 
={}&  \frac{1}{2} \begin{pmatrix}
q_z - \I r_z & q_x - \I r_x - \I q_y - r_y\\
q_x - \I r_x + \I q_y + r_y & -q_z + \I r_z
\end{pmatrix}\;,\label{e_matrix}
\end{align}
\endgroup
\begin{align}
s ={}& \tau_y\otimes\sigma_y = \begin{pmatrix}
0 & 0 & 0 & -1 \\
    0 & 0 & 1 & 0 \\
    0 & 1 &0 & 0 \\
    -1 & 0 & 0 & 0
\end{pmatrix}\;.\label{s_matrix}
\end{align}
Here, the $h_1^n$ are the Fourier components of $h_1(t) = h_1(t+T)$
with
\begin{align}
  h_1(t) ={}& A_x(t)\sigma_x + A_y(t)\sigma_y + A_z(t)\sigma_z =
              \sum_{n=-\infty}^{\infty} e^{-\I n\Omega t} h_1^n\;.
\end{align}
\subsection{L\"owdin high-frequency partitioning}\label{sec:lowd-high-freq}
For a general number of Floquet modes $N$, this truncated Floquet-BdG
Hamiltonian can be only treated numerically. However, in the
off-resonant regime, i.e., when the driving energy $\hbar\Omega$ exceeds the
band width $W$, one can derive an effective BdG Hamiltonian by folding down
higher replica onto the central mode. The suitable tool to accomplish this is L\"owdin partitioning which is 
described in great detail in~[\onlinecite{BirPikusBook,winklerbook}]. In
short: Assume that a Hamiltonian $\mathcal{H}$ can be expressed
as a sum of a Hamiltonian $\mathcal{H}^0$ with known eigenvalues $E_n$
and eigenfunctions $\ket{\psi_n}$ and $\mathcal{H'}$. The latter Hamiltonian is treated
as a perturbation. Decompose $\mathcal{H'}$ further as a sum of a
block diagonal matrix $\mathcal{H}^1$, with subsets $A$ and $B$, and
$\mathcal{H}^2$. Thus, we can write
\begin{align}
  \mathcal{H}=\mathcal{H}^0+\mathcal{H'}=\mathcal{H}^0+\mathcal{H}^1+\mathcal{H}^2.
\end{align}
The goal is to approximate the system consisting of block $A$, $B$ and their
couplings with an effective block $A$ where the effect of block $B$
has been ``folded onto it''. Following~\cite{BirPikusBook,winklerbook}, we define the indices $m, m', m''$
which correspond to the states in set $A$, and the indices $l,l',l''$
to the one of set $B$. The subsets $A$ and $B$ may have degeneracies
but it is crucial that $E_m \neq E_l$. The matrix elements are defined by
\begin{align}
\mathcal{H}'_{ij} :=\mel{\psi_i}{\mathcal{H}'}{\psi_j}.
\end{align}
It can be shown that a non-block-diagonal, anti-Hermitian matrix $S$ exists which
transforms $\mathcal{H}$ into a block diagonal Hamiltonian
$\widetilde{\mathcal{H}}=e^{-S}\mathcal{H} e^{S}$, i.e., matrix $S$
removes the coupling between block $A$ and $B$. This anti-Hermitian matrix can
be approximated in a successive manner which allows for an
approximation of $\widetilde{\mathcal{H}}$,
\begin{align}
\widetilde{\mathcal{H}}={}&\mathcal{H}^{(0)}+\mathcal{H}^{(1)}+\mathcal{H}^{(2)}+\mathcal{H}^{(3)}+\ldots\;.
\label{perturbationformula}
\end{align}
The corresponding terms with indices belonging to the block A are
given by
\begin{widetext}
  \begin{align}
    \mathcal{H}^{(0)}_{m m'} ={}& \mathcal{H}^0_{m m'}\, ,& \label{perturbationformula0}\\
    \mathcal{H}^{(1)}_{m m'} ={}& \mathcal{H}'_{m m'}\, ,\label{perturbationformula1}\\
    \mathcal{H}^{(2)}_{m m'} ={}& \frac{1}{2}\sum_{l}\mathcal{H}'_{m l}\mathcal{H}'_{l m'}\left[\frac{1}{E_m-E_l}+\frac{1}{E_{m'}-E_l}\right],&\label{perturbationformula2}\\
    \mathcal{H}^{(3)}_{m m'} ={}&-\frac{1}{2}\sum_{l, m''}
                                  \left[
                                  \frac{\mathcal{H}'_{m l}\mathcal{H}'_{l m''}\mathcal{H}'_{m'' m'}}
                                  {(E_{m'}-E_l)(E_{m''}-E_l)} + \frac{\mathcal{H}'_{m m''}\mathcal{H}'_{m'' l}\mathcal{H}'_{l m'}}
                                  {(E_{m}-E_l)(E_{m''}-E_l)}
                                  \right]&\notag\\
    {}&+\frac{1}{2}\sum_{l, l'}
        \mathcal{H}'_{m l}\mathcal{H}'_{l l'}\mathcal{H}'_{l' m'}
        \left[
        \frac{1}
        {(E_{m}-E_l)(E_{m}-E_{l'})} + \frac{1}
        {(E_{m'}-E_l)(E_{m'}-E_{l'})}
        \right]\;.\label{perturbationformula3}
  \end{align}
\end{widetext}
In the following we apply the L\"owdin partitioning to the Floquet-BdG
matrix~\eqref{FullFloquetH}. We start by only taking the
central, $n=0$, and the $n=\pm 1$ Floquet
modes into account. Thereby, we fold the $n=\pm 1$ Floquet blocks
onto the central one. 
The L\"owdin correction up to first order in $W/\hbar\Omega$, $W$ the
bandwidth, is then given by
\begingroup
\allowdisplaybreaks{}
\begin{align}
  H_F^{(2)} ={}& \frac{1}{\hbar\Omega} \begin{pmatrix}
[h_1^{-1},h_1^1] & \vb{0}_{2\times 2}\\
\vb{0}_{2\times 2} & -{[h_1^{-1},h_1^1]}^*
  \end{pmatrix}\nonumber\\
  ={}& \frac{q_y r_x - q_x r_y}{\hbar\Omega}\tau_z\otimes\sigma_z +
     \frac{q_x r_z - q_z r_x}{\hbar\Omega}\id_2\otimes\sigma_y\nonumber\\
{}& + \frac{q_z r_y - q_y r_z}{\hbar\Omega}\tau_z\otimes\sigma_x\;.\label{effField}
\end{align}
\endgroup
The second order correction is given by
\begin{align}
  H_F^{(3)} ={}& C_{01} \ \id_2 \otimes \sigma_x + C_{32} \
              \tau_z\otimes \sigma_y + C_{03} \ \id_2 \otimes \sigma_z
\end{align}
with
\begin{align}
  C_{01} ={}& -\frac{1}{\hbar^2\Omega^2} \vb{k}\cdot \begin{pmatrix}
\alpha_x(q_x q_y + r_x r_y)\\
\alpha_y(q_y^2 + q_z^2 + r_y^2 + r_z^2)
  \end{pmatrix}\;,\\
  C_{32} ={}& \ \phantom{-}\frac{1}{\hbar^2\Omega^2} \vb{k}\cdot \begin{pmatrix}
\alpha_x(q_x^2 + q_z^2 + r_x^2 + r_z^2)\\
\alpha_y(q_x q_y + r_x r_y)
  \end{pmatrix}\;,\\
  C_{03} ={}& \ \phantom{-}\frac{1}{\hbar^2\Omega^2} \vb{k}\cdot \begin{pmatrix}
-\alpha_x(q_y q_z + r_y r_z)\\
\alpha_y(q_x q_z + r_x r_z)
  \end{pmatrix}\;.
\end{align}
The terms with the coefficients $C_{01}$ and $C_{32}$ renormalize the
parts in $H_0$ which are due to SOC.\@ The new term proportional to
$\id_2 \otimes \sigma_z$ vanishes in case of an in-plane magnetic
field.  In the following we neglect the second order correction,
leaving us with the effective $4 \times 4$ BdG Floquet Hamiltonian
\begin{align}
  H_{\text{eff,Z}} = H_F^{(0)} + H_F^{(2)}\;.
\end{align} 

\subsection{Spectrum of the effective Hamiltonian}%
\label{subsec:specHeff}
If the driving consists of an in plane magnetic field only the term proportional
to $\tau_z\otimes\sigma_z$ survives in $H_{\text{eff,Z}}$. Then we can
diagonalize the spectrum in an easy way, i.e., without explicitly
using Ferrari's, Descartes' or Euler's solution for quartic functions.
Assuming this parameter setting, the effective static Hamiltonian has
the following form
\begin{widetext}
  \begin{align}
    H_{\text{eff,Z}} ={}&  \left(
                          \begin{array}{cccc}
                            \frac{\hbar^2k^2}{2m^*} - \mu + h_{\text{Z}} & \alpha_y k_y +
                                                                           \I\alpha_x k_x & 0 &
                                                                                                - \Delta \\
                            \alpha_y k_y - \I\alpha_x k_x & \frac{\hbar^2k^2}{2m^*} - \mu - h_{\text{Z}} & \Delta & 0 \\
                            0 & \Delta & -\frac{\hbar^2k^2}{2m^*} + \mu - h_{\text{Z}} &
                                                                                         \alpha_y
                                                                                         k_y -
                                                                                         \I\alpha_x k_x \\
                            -\Delta & 0 & \alpha_y k_y + \I\alpha_x k_x & -\frac{\hbar^2k^2}{2m^*}+\mu+h_{\text{Z}} 
                          \end{array}
                                                                          \right)%
                                                                          \label{hf8}
  \end{align}
\end{widetext}
where $h_{\text{Z}}$, which appears due to the effective field, Eq.~\eqref{effField}, is given by
\begin{align}
  h_{\text{Z}} :={}& \frac{\Lambda_{z}}{\hbar\Omega} = \frac{q_y
                    r_x - q_x r_y}{\hbar\Omega}\;.
\end{align}
The diagonal $2\times 2$-blocks are diagonalized by the matrix
\begin{align}
  V_1(\vb{k})=\left(
  \begin{array}{cc}
    v_1(\vb{k}) & \vb{0}_{2\times 2} \\
    \vb{0}_{2\times 2} & v_2(\vb{k})
  \end{array}
  \right)%
  \label{hf9}
\end{align}
where
\begin{align}
  v_1(\vb{k}) ={}&  \left(
  \begin{array}{cc}
    \gamma_+(\vb{k}) & -\gamma_-(\vb{k})e^{-\I\varphi(\vb{k})} \\
    \gamma_-(\vb{k})e^{\I\varphi(\vb{k})} & \gamma_+(\vb{k})
  \end{array}
  \right)\,,\label{hf10}\\
  v_2(\vb{k}) ={}&  \left(%
  \begin{array}{cc}
    \gamma_-(\vb{k})e^{\I\varphi(\vb{k})} & \gamma_+(\vb{k}) \\
    \gamma_+(\vb{k}) & -\gamma_-(\vb{k})e^{-\I\varphi(\vb{k})}
  \end{array}
  \right)\label{hf11}
\end{align}
with 
\begin{align}
  \gamma_{\pm}(\vb{k})
  =\sqrt{\frac{1}{2}\left(1\pm\frac{h_{\text{Z}}}
    {\sqrt{{\qty(\alpha_x k_x)}^2 + {\qty(\alpha_y k_y)}^2 + h_{\text{Z}}^2}}\right)}\,,
  \label{hf12}
\end{align}
and
\begin{align}
  e^{\I\varphi(\vb{k})}=\frac{\alpha_y k_y - \I\alpha_x k_x}
  {\sqrt{{\qty(\alpha_y k_y)}^2 + {\qty(\alpha_x k_x)}^2}}\,.
  \label{hf13}
\end{align}
The Hamiltonian (\ref{hf8}) is rendered as
\begin{align}
  V^\dagger_1HV_1=\left(
  \begin{array}{cccc}
    \eta_+ & 0 & -\Delta_2 & \Delta^{\ast}_1 \\
    0 & \eta_- & \Delta_1 & \Delta_2 \\
    -\Delta_2 & \Delta^{\ast}_1 & -\eta_- & 0 \\
    \Delta_1 & \Delta_2 & 0 & -\eta_+ 
  \end{array}
  \right)%
  \label{hf14}
\end{align}
where now
\begin{align}
  \eta_{\pm}(\vb{k})=\frac{\hbar^2k^2}{2m^*}-\mu
  \pm\sqrt{{\qty(\alpha_x k_x)}^2 + {\qty(\alpha_y k_y)}^2 + h_{\text{Z}}^2}\;.
  \label{hf15}
\end{align}
Explicitly, we find that 
\begin{align}
  \Delta_1=2\Delta\gamma_+\gamma_-e^{i\varphi}
  \ ,\quad\
  \Delta_2=\Delta(\gamma_+^2-\gamma_-^2)\,,
  \label{hf16}
\end{align} 
and hence the intraband
pairing $\Delta_1$ is an odd function in the wave vector $\vb{k}$. In contrast, the
interband pairing $\Delta_2$ is an even function of momentum.  The
transformed Hamiltonian Eq.~\eqref{hf14} thus reveals the presence of an
effective $p$-wave intravalley pairing, a necessary requirement for
topological superconductivity~\cite{Sato_2017}. To fully uncover the topological properties
of the effective Hamiltonian, we now follow Ref.~\onlinecite{Marganska_2018} and apply the
transformation
\begin{align}
  V_2=\left(
  \begin{array}{cccc}
    \delta_+ & 0 & \delta_- & 0\\
    0 & \delta_- & 0 & \delta_+\\
    -\delta_- & 0 & \delta_+ & 0\\
    0 & -\delta_+ & 0 & \delta_-
  \end{array}
  \right)%
  \label{hf17}
\end{align}
with 
\begin{align}
  \delta_{\pm}
  =\sqrt{\frac{1}{2}\left(1 \pm \frac{(\eta_+ + \eta_-)/2}
    {\sqrt{{\qty(\eta_++\eta_-)}^2/4 + \Delta^2_2}}\right)}\;.
  \label{hf18}
\end{align}
This transformation results in
\begin{align}
  V^\dagger_2V^\dagger_1HV_1V_2 = \left(
  \begin{array}{cccc}
    \lambda_+ & -\Delta^{\ast}_1 & 0 & 0\\
    -\Delta_1 & -\lambda_+ & 0 & 0\\
    0 & 0 & \lambda_- & \Delta^{\ast}_1\\
    0 & 0 & \Delta_1 & -\lambda_-
  \end{array}
  \right)%
  \label{hf19}
\end{align}
where
\begin{align}
  \lambda_{\pm}=\frac{\eta_+ - \eta_-}{2}
  \pm\sqrt{\frac{{\qty(\eta_+ + \eta_-)}^2}{4} + \Delta^2_2}\,.
  \label{hf20}
\end{align}
Finally we can diagonalize the remaining $2\times 2$ matrices. The four eigenvalues are given by
\begin{align}
  \lambda_{mn} ={}&  m (h_{\text{Z},c}^2 + h_{\text{Z}}^2 + \ka^2(\alphaT^2 +
  \betaT^2 - 2\hop\mu + 2\alphaT\betaT\cos(2\theta))\nonumber\\
{}& + \hop^2\ka^4 + 2 n\sqrt{\mathcal{W}})^{\frac{1}{2}}\label{eigenvals}
\end{align}
with $m=\pm 1$, $n=\pm 1$ and the critical field 
\begin{align}
  |h_{\text{Z},c}| = \sqrt{\Delta^2 + \mu^2}\;.\label{h_criticalAppendix}
\end{align}
We introduced the abbreviation 
\begin{align}
  \mathcal{W} :={}& h_{\text{Z}}^2 h_{\text{Z},c}^2 + \ka^2\mu((\alphaT^2 + \betaT^2)\mu - 2h_{\text{Z}}^2\hop +
  2\alphaT\betaT\mu\cos(2\theta))\nonumber\\ 
{}& + \ka^4\hop(h_{\text{Z}}^2\hop- 2(\alphaT^2 + \betaT^2)\mu
  - 4\alphaT\betaT\mu\cos(2\theta))\nonumber\\ 
{}& + \ka^6\hop^2(\alphaT^2 + \betaT^2 + 2\alphaT\betaT\cos(2\theta))\;,
\end{align}
where we used polar coordinates with $\ka_x = \ka\cos(\theta)$, $\ka_y =
\ka\sin(\theta)$, and defined dimensionless quantities $\ka:= ka$, $\alphaT:=\alpha/a$, $\betaT:=
\beta/a$, with $a$ the lattice constant. Finally, $\hop
:=\hbar^2/(2m^*a^2)$ is the hopping energy. 
\section{Floquet BdG Hamiltonian in the tight-binding formulation}\label{app:FloquetTB}
\subsection{Tight-binding version of the static Hamiltonian}
For many numerical purposes it is convenient to work with a tight-binding
Hamiltonian which reduces to the static Hamiltonian Eq.~\eqref{general2}
in the long wave length limit. Working on a square lattice with spacing $a$, the latter quantity is
reformulated as
\begin{align}
{}&  \CH_0 + \CH_{\Delta} =\nonumber\\
{}&                            \sum_{ml\sigma}\Big[(4\hop-\mu) c^\dagger_{ml\sigma}
                            c_{ml\sigma}
    -\hop(c^\dagger_{(m+1)l\sigma}c_{ml\sigma}\nonumber\\
{}& \phantom{\sum_{ml\sigma}}\, +
                            c^\dagger_{m(l+1)\sigma} c_{ml\sigma}+{\rm H.c.})\Big]
  \nonumber\\
  {}& \, + \sum_{ml}\Bigl[-\frac{\alphaT_x}{2}
    (c^\dagger_{(m+1)l\uparrow} c_{ml\downarrow} -
      c^\dagger_{ml\uparrow} c_{(m+1)l\downarrow} + {\rm H.c.})
    \nonumber\\
   {}& \qquad\, + i\frac{\alphaT_y}{2}
    (c^\dagger_{m(l+1)\uparrow} c_{ml\downarrow} -
       c^\dagger_{ml\uparrow} c_{m(l+1)\downarrow}-{\rm H.c.})
    \Bigr]\nonumber\\
  {}& \,-\frac{\Delta}{2}\sum_{ml}
  \left(c^\dagger_{ml\uparrow} c^\dagger_{ml\downarrow} -
      c^\dagger_{ml\downarrow} c^\dagger_{ml\uparrow}
    +{\rm H.c.}\right)\,,
  \label{tb1}
\end{align}
where the operators $c^\dagger_{ml\sigma}$, $c_{ml\sigma}$ create and annihilate,
respectively, a particle with spin $\sigma\in\{\uparrow,\downarrow\}$ at
lattice site $\vb{r}=ma\,\vb{e}_x + la\,\vb{e}_y$. We have redefined the
spin-orbit parameters as $\alphaT_x:=\alpha_x/a$,
$\alphaT_y:=\alpha_y/a$.

Let us now concentrate on an infinite stripe along the $y$-direction with
$L$ transversal lattice sites enumbered by $m\in\{0,\dots,L-1\}$.
Introducing the operators
\begin{align}
  c^\dagger_{m\sigma}(k) = \sqrt{\frac{a}{2\pi}}\sum_l e^{\I ka l}c^\dagger_{ml\sigma}
  \label{tb2}
\end{align}
for each wave number $k\in[-\pi/a,\pi/a]$ along with the Nambu spinor
\begin{align}
  \Psi^\dagger_q(k) ={}& (\phi^\dagger_0(k),\dots,\phi^\dagger_{L-1}(k))\;,\quad
  q\in\{0,\dots,4L-1\}\;,
  \label{tb3}\\
\intertext{where}
  \phi^\dagger_m(k) ={}&  (c^\dagger_{m\uparrow}(k),c^\dagger_{m\downarrow}(k),
  c_{m\uparrow}(-k),c_{m\downarrow}(-k))\,,
  \label{tb4}
\end{align}
the static Hamiltonian can be expressed as
\begin{align}
  \CH_0 + \CH_{\Delta} = \frac{1}{2}\int_{-\pi/a}^{\pi/a}dk\sum_{q,q^{\prime}=0}^{4L-1}
  \Psi^\dagger_q(k)H_{qq^{\prime}}(k)\Psi_{q^{\prime}}(k)\,.
  \label{tb5}
\end{align}
Here the BdG Hamiltonian in stripe geometry reads
\begin{align}
  H_{qq^{\prime}}(k)=\left(
  \begin{array}{ccccc}
    a(k) & b & & & \\
    b^\dagger & a(k) & b & & \\
    & b^\dagger & a(k) & b & \\
    & & \ddots & \ddots & \ddots  
  \end{array}
  \right)\;,\label{tb6}
\end{align}
with
\begin{align}
{}&  a(k) = \left(
  \begin{array}{cccc}
    \eta(k) & \alphaT_y\sin(ka) & 0 & -\Delta \\
    \alphaT_y\sin(ka) & \eta(k) & \Delta & 0 \\
    0 & \Delta & -\eta(k) & \alphaT_y\sin(ka) \\
    -\Delta & 0 & \alphaT_y\sin(ka) & -\eta(k) 
  \end{array}
  \right)\;,\\
{}& \eta(k) = (4 - 2\cos(ka))\hop - \mu\,,\label{tb7}
\end{align}
and
\begin{align}
  b=\left(
  \begin{array}{cccc}
    -\hop & \alphaT_x/2 & 0 & 0 \\
    -\alphaT_x/2 & -\hop & 0 & 0 \\
    0 & 0 & \hop & -\alphaT_x/2 \\
    0 & 0 & \alphaT_x/2 & \hop 
  \end{array}
  \right)\,.\label{tb9}
\end{align}
Note that the matrix (\ref{tb6}) is real and symmetric.

For a stripe along the $x$-direction with again
$L$ transversal lattice sites enumbered now by $l\in\{0,\dots,L-1\}$ one
defines operators
\begin{align}
  c^\dagger_{l\sigma}(k) = \sqrt{\frac{a}{2\pi}}\sum_m e^{\I ka m} c^\dagger_{ml\sigma}\,,
  \label{tb10}
\end{align}
and the entries of the spinor (\ref{tb3}) now read
\begin{align}
  \phi^\dagger_m(k) = (c^\dagger_{l\uparrow}(k),c^\dagger_{l\downarrow}(k),
  c_{l\uparrow}(-k),c_{l\downarrow}(-k))\,.
  \label{tb11}
\end{align}
The resulting BdG matrix is again of the form (\ref{tb6})
where now
\begin{align}
{}&  a(k) =\nonumber\\ 
{}&\left(
  \begin{array}{cccc}
    \eta(k) & \I\alphaT_x\sin(ka) & 0 & -\Delta \\
    -\I\alphaT_x\sin(ka) & \eta(k) & \Delta & 0 \\
    0 & \Delta & -\eta(k) & -\I\alphaT_x\sin(ka) \\
    -\Delta & 0 & i\alphaT_x\sin(ka) & -\eta(k) 
  \end{array}
  \right)\,,\label{tb12}
\end{align}
\begin{align}
  b = \left(
  \begin{array}{cccc}
    -\hop & -\I\alphaT_y/2 & 0 & 0 \\
    -\I\alphaT_y/2 & -\hop & 0 & 0 \\
    0 & 0 & \hop & -\I\alphaT_y/2 \\
    0 & 0 & -\I\alphaT_y/2 & \hop 
  \end{array}
  \right)\,.\label{tb13}
\end{align}
\subsection{Tight-binding Floquet-BdG Hamiltonian in Nambu-Sambe space}
Again, we consider the same driving as in Eq.~\eqref{driving}. Taking
into account $2N+1$ Fourier modes with labels
$n\in\{-N,-N+1,\dots,N\}$ around the central temporal Brillouin zone,
the Nambu-Sambe spinor has $4L(2N+1)$ components. The Floquet-BdG
Hamiltonian takes the form
\begin{align}
  H_F(k) = H_{\rm btd}(k)+H_{\Delta}\;,
  \label{periodic4}
\end{align}
where the first contribution is block-tridiagonal,
\begin{widetext}
  \begin{align}
    H_{\rm btd}(k) = \left(
    \begin{array}{ccccc}
      H(k)+N\hbar\Omega T^z & D & & & \\
      D^+ & H(k)+(N-1)\hbar\Omega T^z & D & & \\
                            & \ddots & \ddots & \ddots & \\
                            & & & & D\\
                            & & & D^+ & H(k)-N\hbar\Omega T^z\\
    \end{array}
    \right)%
    \label{periodic5}
  \end{align}
\end{widetext}
with $H(k)$ being the $4L\times4L$ BdG matrix (\ref{tb6}), but now without the
superconducting coupling so that its diagonal blocks read
\begin{align}
  a(k) = \left(
  \begin{array}{cccc}
    \eta(k) & \alphaT_y\sin(ka) & 0 & 0 \\
    \alphaT_y\sin(ka) & \eta(k) & 0 & 0 \\
    0 & 0 & -\eta(k) & \alphaT_y\sin(ka) \\
    0 & 0 & \alphaT_y\sin(ka) & -\eta(k) 
  \end{array}
  \right)\,.%
  \label{periodic6}
\end{align}
The matrix $T^z$ is diagonal,
\begin{align}
  T^z = \left(
  \begin{array}{cccc}
    \tau^z\otimes\id_2 &  & & \\
    & \tau^z\otimes\id_2 & & \\
    & & \ddots & \\
    & & & \tau^z\otimes\id_2 \\
  \end{array}
  \right)\;.%
  \label{periodic7}
\end{align}
The driving is further implemented in the block-diagonal matrix
\begin{align}
  D=\left(
  \begin{array}{cccc}
    d &  & & \\
    & d & & \\
    & & \ddots & \\
    & & & d \\
  \end{array}
  \right)\;,%
  \label{periodic9}
\end{align}
with $d$ defined in Eq.~\eqref{d_matrix}.
Finally, the superconducting coupling is included in the contribution
\begin{align}
  H_{\Delta}=\left(
  \begin{array}{cccc}
     & & & S \\
    & & \reflectbox{$\ddots$} & \\
    & S & & \\
    S & & &  \\
  \end{array}
  \right)%
  \label{periodic11}
\end{align}
where the $4L\times 4L$ matrix $S$ is given by
\begin{align}
  S=\left(
  \begin{array}{cccc}
    s &  & & \\
    & s & & \\
    & & \ddots & \\
    & & & s \\
  \end{array} 
  \right)\;,\label{periodic12}
\end{align}
with $s$ defined in Eq.~\eqref{s_matrix}.

As mentioned in Sec.~\ref{sec:bdg-hamilt-sambe}, the correct
anti-diagonal position of the superconducting gap $\Delta$ in the
Floquet space, connecting the $n$ with the $-n$ Floquet mode, results
in a quasienergy spectrum which shows the usual appearance of
replica. The spectrum of the Floquet Hamiltonian in stripe geometry is
shown in Fig.~\ref{fig:replica}. We show the central Floquet band and the two
Floquet replicas above it.
\subsection{Tight-binding spectra and Chern numbers}\label{sec:devi-from-effect}
\begin{figure}[t]
\centering
\includegraphics[width=\linewidth]{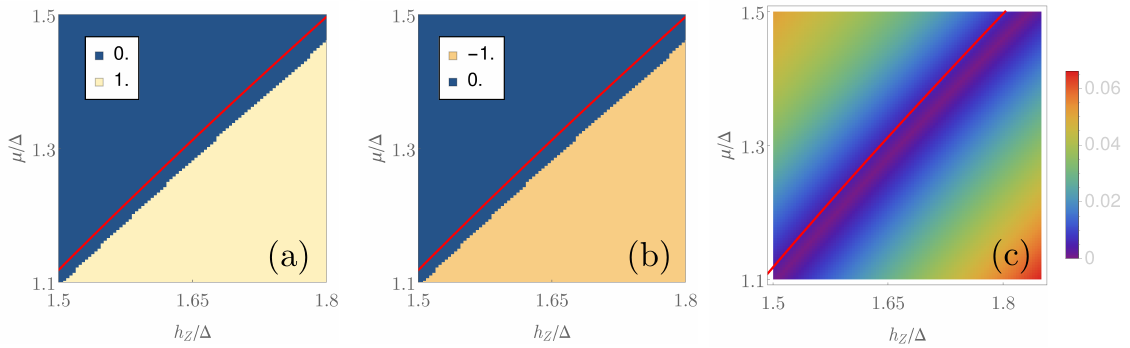}
\caption{Chern numbers of the second (a) and third (b) band of the
  first Floquet Brillouin zone of $H_F$. (c) Energy gap between the second and third band of the central
  Floquet mode of $H_F$ in units of $\hop$. Values are calculated
  using the tight-binding formulation and shown for different field strengths $h_{\text{Z}}$ and
chemical potential $\mu$ in units of the superconducting gap $\Delta$. The red line
indicates $|h_{\text{Z},c}|$, Eq.~\eqref{h_criticalAppendix}. Parameters 
used: $N_F=5$  and
the frequency is $\hbar\Omega/\hop = 20$, $q_x=q_z=0$, $r_z=1\ \hop$, $\alpha
= 0.1\ \hop a$, $\beta = 0.37\ \hop a$.}\label{fig:ChernZoom} 
\end{figure}

To determine from the 2D spectrum whether or not we can find
topologically protected boundary modes in a stripe geometry, we
calculate the Chern numbers $C_\alpha$ for the bulk system. The
latter are determined by integrating over the Berry curvature $\vb{F}_\alpha(\vb{k})$~\cite{Thouless1982a,Berry1984,Simon1983},
\begin{align}\label{Cherndef}
C_\alpha ={}& \frac{1}{2\pi}\int_{\text{BZ}}d^2k\,\vb{F}_\alpha(\vb{k})\cdot{\hat{\vb{z}}}\;,
\end{align}
with 
\begin{align}
  \vb{F}_\alpha(\vb{k}) ={}& \sum_{\beta \neq \alpha}\mathrm{Im}
                          \frac{\mel{u_{\vb{k}\alpha}^n}{\vb{\nabla}_{\vb{k}}H_F}{u_{\vb{k}\beta}^n}
                          \times
                          \mel{u_{\vb{k}\beta}^n}{\vb{\nabla}_{\vb{k}}H_F}{u_{\vb{k}\alpha}^n}}{{(\varepsilon_{\vb{k}\alpha}
                          - \varepsilon_{\vb{k}\beta})}^2}\;.
\end{align}
The $\ket{u_{\vb{k}\alpha}^n}$, $\alpha=1,2,3,4$, are the eigenstates
of the $n$-th Floquet mode of the Floquet-BdG tight-binding Hamiltonian.
In Figs.~\ref{fig:ChernZoom}(a) and  \ref{fig:ChernZoom}(b) we also show the Chern number of
the second and third Floquet bands ($\alpha = 2,3$) within the first Floquet Brillouin
zone ($n=0$). We restricted to a smaller range of chemical potentials compared
to Fig.~\ref{fig:toptrans4}. In this way we notice a small
difference between the phase boundary as evinced from the numerically
calculated Chern numbers, and the one obtained from the effective
continuum Hamiltonian, Eq.~\eqref{h_criticalAppendix}, given by the solid red
line. This difference is due to the fact that the gap closing
condition is slightly shifted compared to the tight-binding model, as
shown in Fig.~\ref{fig:ChernZoom}(c).

\begin{figure}[t]
\centering
\includegraphics[width=\linewidth]{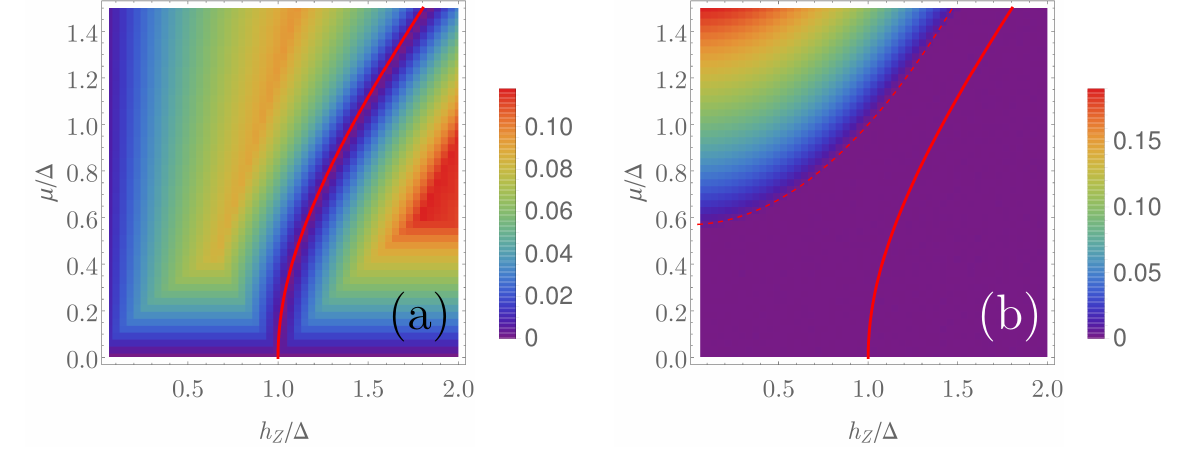}
\caption{Smallest energy gap in units of $\hop$. (a) Shown is the gap
  between the second and third band of
  the central Floquet mode of $H_F$. (b) Smallest energy gap between different Floquet replica, of $H_F$ for different
  values of field strength $h_{\text{Z}}$ and chemical potential
  $\mu$. Parameters used: $N_F=5$ and the frequency is
  $\hbar\Omega/\hop = 16$, $q_x=q_z=0$, $r_z=1\ \hop$, $\alpha = 0.1\ \hop a$,
  $\beta = 0.37\ \hop a$.}\label{fig:pigapomega16}
\end{figure}

Finally, we show in Fig.~\ref{fig:pigapomega16} energy gaps for a lower
driving frequency $\hbar\Omega = 16\hop$. In Fig.~\ref{fig:pigapomega16}(a) the smallest energy
gap between the bands of the central Floquet mode is shown on a larger
parameter scale with the red line indicating the analytically found
phase boundary. As one can see from Fig.~\ref{fig:pigapomega16}(b),
which shows the smallest energy gap between different Floquet replica,
for the largest part of the parameter space the driving frequency is
too low to be in the off-resonant regime. The dashed line indicates
the parameters where the modes from two different replica touch.
%
\begin{figure}[t] 
  \centering
  \includegraphics[width=0.5\linewidth]{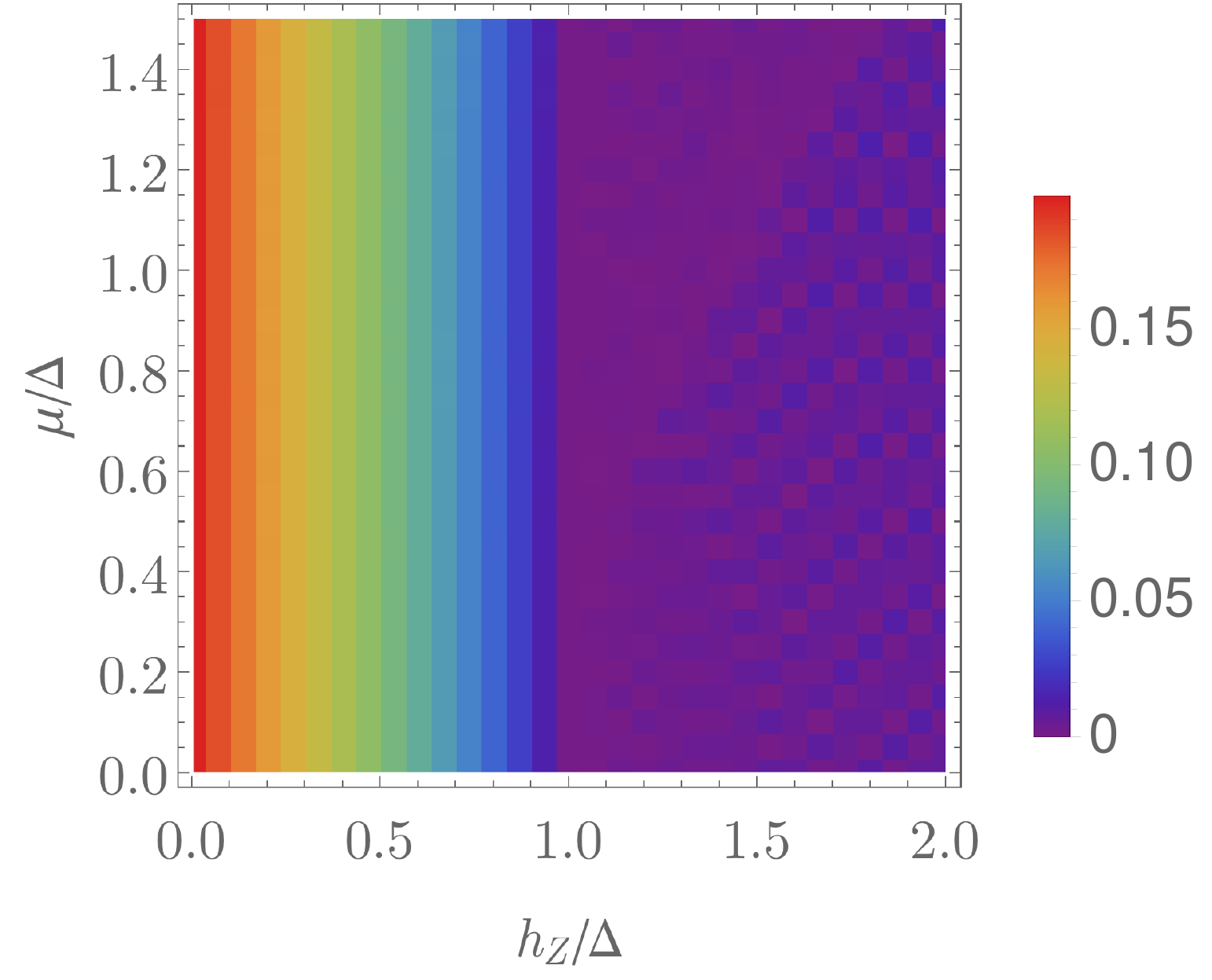}
  \caption{Central gap of the 2D Floquet spectrum in units of $\hop$
    for $\alpha=\beta$. The gap closes for $h_{\text{Z}} =
    \sqrt{\Delta^2 + \mu^2}$ at
    $k=0$. However, there is already a gap closing at finite wave
    vectors for $h_{\text{Z}} \geq \Delta$, see discussion in Sec.~\ref{sub:alphabeta}. The
    parameters are $\alpha=\beta=0.37\ \hop a$, $N_F=3$,
    $\Delta=0.1\ \hop$, $r_x=1\ \hop$, $r_y=0.2\ \hop$,
    $r_z=q_x=q_y=0$.}%
\label{fig:centralgapAlphaeqBeta}
\end{figure}
%
\begin{figure}[t]
  \centering
  \includegraphics[width=\linewidth]{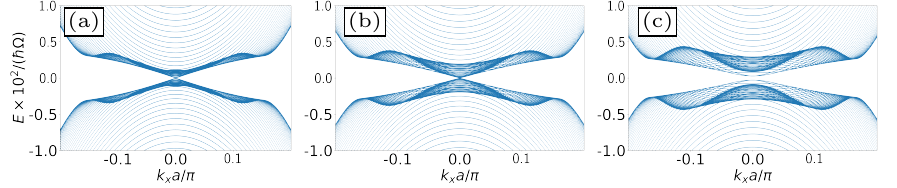}
  \caption{Quasienergy spectrum of $H_F(k)$ in stripe geometry with a
    width of $N_y=200$ transverse channels and for SOC parameters
    $\alpha = \beta$. The three panels show different field
    strengths. The field is changed via $q_y$ (for fixed frequency
    $\hbar\Omega=20\ \hop$
    and $N_F=4$) for $\alpha=\beta=0.1\ \hop a$. $[\Delta=0.1\ \hop,\quad
    \mu=0.1\; \hop,\quad r_x=2\; \hop,\quad r_y=0.2\ \hop,\quad q_x=0,\quad
    h_{\text{Z},c}/\Delta =\sqrt{2}]$: (a)
    $q_y=1.2\; \hop$, thus $h_{\text{Z}}/h_{\text{Z},c} = 0.85$, (b) $q_y=1.4\; \hop$, thus $h_{\text{Z}}/h_{\text{Z},c} = 0.99$, (c) $q_y=1.6\; \hop$, thus $h_{\text{Z}}/h_{\text{Z},c} = 1.13$.}%
  \label{fig:rx1}
\end{figure} 
\begin{figure}[t]
  \centering
  \includegraphics[width=\linewidth]{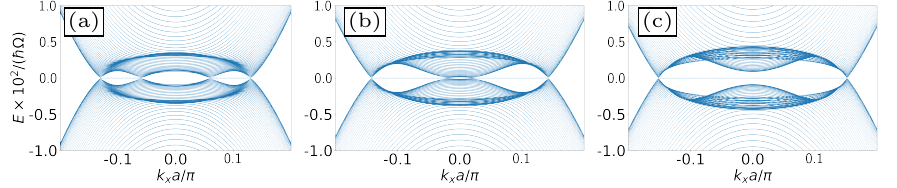} 
  \caption{Quasienergy spectrum of $H_F(k)$ in stripe geometry with a
    width of $N_y=200$ transverse channels and for SOC parameters
    $\alpha = -\beta$. The three panels show different field
    strengths. The field is changed via $q_y$ (for fixed frequency
    $\Omega=20\ \hop$ and $N_F=4$) for $\alpha=-\beta=0.1\ \hop
    a$.
    $[\Delta=0.1\ \hop,\ \mu=0.1\ \hop,\quad r_x=2\ \hop,\quad
    r_y=0.2\ \hop,\quad q_x=0,\quad h_{\text{Z},c}/\Delta =\sqrt{2}]$: (a) $q_y=1.2\; \hop$, thus
    $h_{\text{Z}}/h_{\text{Z},c} = 0.85$, (b) $q_y=1.4\; \hop$, thus $h_{\text{Z}}/h_{\text{Z},c} = 0.99$,
    (c) $q_y=1.6\; \hop$, thus $h_{\text{Z}}/h_{\text{Z},c} = 1.13$.}%
  \label{fig:rx2}
\end{figure}
\section{Characterization of the topological phase and edge states}%
\label{sec:char-topol-phase}
We are going to discuss different conditions for the Rashba
and Dresselhaus SOC parameters. The ratio between both controls the
appearance of flat bands. 
From Eq.~\eqref{eigenvals}, we can extract the field $h_{\text{Z}}$ at which the
central gap closes. One finds the following four conditions,
\begin{align}
{}&  h_{Z,mn}(\ka) =\nonumber\\ 
{}& m \sqrt{\hop^2\ka^4 + h_{\text{Z},c}^2 + n\sqrt{-2\ka^2\Delta^2
                  \gamma} - \frac{\ka^2}{2}(4\hop\mu + \gamma)}\;,\label{conditionhzk}
\end{align}
with $m=\pm 1$, $n=\pm 1$ and
\begin{align}
  \gamma := 2(\alphaT^2 + \betaT^2 + 2\alphaT\betaT\cos(2\theta))\;.
\end{align}
We distinguish three cases:
\subsection{\texorpdfstring{$\abs{\alpha}\neq\abs{\beta}$}{|alpha|!=|beta|}}
A gap closing only happens for $\ka=0$; thus one ends up with
the critical fields $h_{\text{Z},\pm,n}(\ka=0)\equiv h_{\text{Z}\pm}(\ka=0) := \pm h_{\text{Z},c} =
\pm \sqrt{\Delta^2 + \mu^2}$. An example for the
$\abs{\alpha}\neq\abs{\beta}$ condition is shown in
Fig.~\ref{fig:toptrans4}. This case can be well
understood in terms of Chern numbers.\newline

As one can see, the gap closing at $\ka=0$ does not depend on
the SOC.\@ However, in the following it will be shown that a gap also closes at a
finite momentum $\ka=\sqrt{\mu/\hop}$ if $\abs{\alpha} = \abs{\beta}$. The
needed critical field in this case is equal or smaller than $h_{\text{Z},c}$. Let
us first consider the case $\alpha = \beta$.
\subsection{\texorpdfstring{$\alpha = \beta$}{alpha=beta}}\label{sub:alphabeta}
In this case, $\gamma$ simplifies to $\gamma=4\alphaT^2(1 +
\cos(2\theta))$ and a necessary condition for a gap closing at a
finite $\ka$ is $\theta = \pi/2$. In other terms, $\ka_x = 0$ and
$\ka_y = \ka$, i.e., a gap closing can be found along the
$y$-direction in $k$-space. Equation~\eqref{conditionhzk} yields
(sign $n$ is redundant since $\gamma =0$)
\begin{align}
  h_{\text{Z},\pm}(\ka) = \pm h_{\text{Z},c}(\ka) :={}& \sqrt{\Delta^2 + {\qty(\mu - \ka^2 \hop)}^2}\;.\label{hSOC}
\end{align}
Thus, the smallest
field at which the gap closes is
\begin{align}
  \abs{h_{\text{Z},c}(\ka = \mu/\hop)} = \Delta \leq
h_{\text{Z},c}\;.
\end{align}
Since we are free to choose $k$ as large as we want
in Eq.~\eqref{hSOC}, there is no gap re-opening if the field is enlarged, as shown in
Fig.~\ref{fig:centralgapAlphaeqBeta}. This is in contrast to the
$\abs{\alpha} \neq \abs{\beta}$ case. The quasienergy spectrum in case
of a stripe geometry with confinement in the $y$-direction is
exemplarily shown in Fig.~\ref{fig:rx1} for various values of the field strength
changed via $q_y$. As expected, for this orientation of the stripe no
closing occurs.
\subsection{\texorpdfstring{$\alpha=-\beta$}{alpha=-beta}}
Here, the only difference to the $\alpha=\beta$ case is the direction
at which the gap closing happens which is $\theta=0$. As for case~\ref{sub:alphabeta},
depending on the stripe direction, one can see either a gap-closing
and re-opening if increasing the field $h_{\text{Z}}$ or only a
closing without gap re-opening. The energy spectrum for such a case in
stripe geometry is shown in Fig.~\ref{fig:toptrans5} and reported here
also in Fig.~\ref{fig:rx2}. In addition, we
plot in Fig.~\ref{fig:central_gap} the central energy gap as a
function of the longitudinal wave vector and the field
$h_{\text{Z}}$. Figure~\ref{fig:central_gap} shows nicely how the
separated flat bands merge to one with increasing $h_{\text{Z}}$.
\subsection{Partial Berry-phase}\label{sec:partial-berry-phase}%
\begin{figure}[t]
  \centering
  \includegraphics[width=\linewidth]{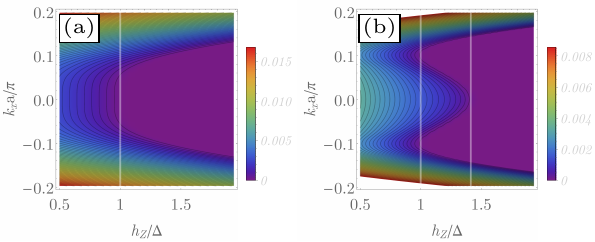}
  \caption{Central gap, in units of $\hbar\Omega$, plotted for $\alpha
    = -\beta = 0.1\ \hop$ with 
    (a) $\mu=0$ and (b) $\mu=0.1\ \hop$. In (a) the vertical line indicates
    the critical field $h_{\text{Z}}/\Delta =1$ below which one finds a gapped
    system. In (b) the second vertical line indicates $h_{\text{Z},c}$. Other parameters: $N_F=2$,
    $\Delta=0.1\ \hop$, $r_x=1\ \hop$, $r_y=0.2\ \hop$,
    $r_z=q_x=q_y=0$.}%
  \label{fig:central_gap}
\end{figure}

From the spectrum analysis above and especially Eq.~\eqref{hSOC}, the
region where we find disconnected flat bands [Fig.~\ref{fig:rx2} (a) and
(b)] is given by $\Delta< h_{\text{Z}} <\abs{h_{\text{Z},c}}$. This
can be understood by counting the band touchings of the two central
bands (for $\mu=0$ and $\mu=0.1\ \hop$ the central gap is plotted in
Fig.~\ref{fig:central_gap}).
From Eq.~\eqref{hSOC}
this happens at
\begin{align}
  \ka_{\pm} ={}& \sqrt{\frac{\mu}{\hop} \pm \frac{1}{\hop}\sqrt{h_{\text{Z}}^2 - \Delta^2}}\;.
\end{align}
For $h_{\text{Z}} \geq h_{\text{Z},c}$, we are left with only one real
value, $\ka_+$.  Since the gap at $\ka=0$ is closed for fields
$h_{\text{Z}}$ larger than $\Delta$ for $\abs{\alpha}=\abs{\beta}$,
one fails to calculate the Chern numbers of the central two
bands. However, examining Fig.~\ref{fig:rx2}, one expects the
existence of a topological quantity which distinguishes the parameter
space where we find disconnected flat bands from the one where we find
only one connected flat band as shown in Fig.~\ref{fig:rx2} (c).  To
connect this change of band-touchings with a change of topology of the
spectrum, we follow the method presented in Refs.~[\onlinecite{Deng2014,Deng_2013}]
and calculate the ``partial'' topological quantum number called
\textit{partial Berry-phase sum parity} (PBSP). According
to Refs.~[\onlinecite{Deng2014,Deng_2013}] we fix one of the wave vector components
by $k_p\in\{0, \pi\}$, $p\in\{x,y\}$ and calculate the PBSP (in the
following we choose $k_x$ to be fixed)
\begin{align}
  P_B ={}&  {(-1)}^{\text{mod}_{2\pi}(B_{+,k_x})/\pi},\quad B_+ = B_{k_x,1} + B_{k_x,2}
\end{align}
with
\begin{align}
  B_{k_x,n} ={}& \I \int_{-\pi}^{\pi} dk_y \,
           \braket{\phi_{n,k_y}}{\partial_{k_y} \phi_{n,k_y}}\;,
\end{align}
where $n$ is the index of the two negative occupied bands. The
numerical evaluation of $B_n$, which has to guarantee numerically gauge
invariance, is described in detail in Appendix D
of Ref.~[\onlinecite{Deng_2013}]. The result for $k_x = 0$, $\alpha_x = 0$ is shown
in Fig.~\ref{fig:toptrans5}(d). For $k_x = \pi/ a$ one finds the same result.
%
\bibliographystyle{apsrev4-1}
\bibliography{pw_bibdatabase_all}

\begin{thebibliography}{46}%
\makeatletter
\providecommand \@ifxundefined [1]{%
 \@ifx{#1\undefined}
}%
\providecommand \@ifnum [1]{%
 \ifnum #1\expandafter \@firstoftwo
 \else \expandafter \@secondoftwo
 \fi
}%
\providecommand \@ifx [1]{%
 \ifx #1\expandafter \@firstoftwo
 \else \expandafter \@secondoftwo
 \fi
}%
\providecommand \natexlab [1]{#1}%
\providecommand \enquote  [1]{``#1''}%
\providecommand \bibnamefont  [1]{#1}%
\providecommand \bibfnamefont [1]{#1}%
\providecommand \citenamefont [1]{#1}%
\providecommand \href@noop [0]{\@secondoftwo}%
\providecommand \href [0]{\begingroup \@sanitize@url \@href}%
\providecommand \@href[1]{\@@startlink{#1}\@@href}%
\providecommand \@@href[1]{\endgroup#1\@@endlink}%
\providecommand \@sanitize@url [0]{\catcode `\\12\catcode `\$12\catcode
  `\&12\catcode `\#12\catcode `\^12\catcode `\_12\catcode `\%12\relax}%
\providecommand \@@startlink[1]{}%
\providecommand \@@endlink[0]{}%
\providecommand \url  [0]{\begingroup\@sanitize@url \@url }%
\providecommand \@url [1]{\endgroup\@href {#1}{\urlprefix }}%
\providecommand \urlprefix  [0]{URL }%
\providecommand \Eprint [0]{\href }%
\providecommand \doibase [0]{http://dx.doi.org/}%
\providecommand \selectlanguage [0]{\@gobble}%
\providecommand \bibinfo  [0]{\@secondoftwo}%
\providecommand \bibfield  [0]{\@secondoftwo}%
\providecommand \translation [1]{[#1]}%
\providecommand \BibitemOpen [0]{}%
\providecommand \bibitemStop [0]{}%
\providecommand \bibitemNoStop [0]{.\EOS\space}%
\providecommand \EOS [0]{\spacefactor3000\relax}%
\providecommand \BibitemShut  [1]{\csname bibitem#1\endcsname}%
\let\auto@bib@innerbib\@empty
\bibitem [{\citenamefont {Oka}\ and\ \citenamefont
  {Kitamura}(2019)}]{Oka_2019}%
  \BibitemOpen
  \bibfield  {author} {\bibinfo {author} {\bibfnamefont {T.}~\bibnamefont
  {Oka}}\ and\ \bibinfo {author} {\bibfnamefont {S.}~\bibnamefont {Kitamura}},\
  }\href {\doibase https://doi.org/10.1146/annurev-conmatphys-031218-013423}
  {\bibfield  {journal} {\bibinfo  {journal} {Annual Review of Condensed Matter
  Physics}\ }\textbf {\bibinfo {volume} {10}},\ \bibinfo {pages} {387}
  (\bibinfo {year} {2019})}\BibitemShut {NoStop}%
\bibitem [{\citenamefont {Shirley}(1965)}]{Shirley1965}%
  \BibitemOpen
  \bibfield  {author} {\bibinfo {author} {\bibfnamefont {J.~H.}\ \bibnamefont
  {Shirley}},\ }\href {\doibase 10.1103/PhysRev.138.B979} {\bibfield  {journal}
  {\bibinfo  {journal} {Physical Review}\ }\textbf {\bibinfo {volume} {138}},\
  \bibinfo {pages} {B979} (\bibinfo {year} {1965})}\BibitemShut {NoStop}%
\bibitem [{\citenamefont {Grifoni}\ and\ \citenamefont
  {H{\"a}nggi}(1998)}]{Grifoni1998}%
  \BibitemOpen
  \bibfield  {author} {\bibinfo {author} {\bibfnamefont {M.}~\bibnamefont
  {Grifoni}}\ and\ \bibinfo {author} {\bibfnamefont {P.}~\bibnamefont
  {H{\"a}nggi}},\ }\href {\doibase 10.1016/S0370-1573(98)00022-2} {\bibfield
  {journal} {\bibinfo  {journal} {Physics Reports}\ }\textbf {\bibinfo {volume}
  {304}},\ \bibinfo {pages} {229} (\bibinfo {year} {1998})}\BibitemShut
  {NoStop}%
\bibitem [{\citenamefont {Bovensiepen}\ and\ \citenamefont
  {Kirchmann}(2012)}]{Bovensiepen_2012}%
  \BibitemOpen
  \bibfield  {author} {\bibinfo {author} {\bibfnamefont {U.}~\bibnamefont
  {Bovensiepen}}\ and\ \bibinfo {author} {\bibfnamefont {P.~S.}\ \bibnamefont
  {Kirchmann}},\ }\href {\doibase https://doi.org/10.1002/lpor.201000035}
  {\bibfield  {journal} {\bibinfo  {journal} {Laser {\&} Photonics Reviews}\
  }\textbf {\bibinfo {volume} {6}},\ \bibinfo {pages} {589} (\bibinfo {year}
  {2012})}\BibitemShut {NoStop}%
\bibitem [{\citenamefont {Wang}\ \emph {et~al.}(2013)\citenamefont {Wang},
  \citenamefont {Steinberg}, \citenamefont {Jarillo-Herrero},\ and\
  \citenamefont {Gedik}}]{10.1126/science.1239834}%
  \BibitemOpen
  \bibfield  {author} {\bibinfo {author} {\bibfnamefont {Y.~H.}\ \bibnamefont
  {Wang}}, \bibinfo {author} {\bibfnamefont {H.}~\bibnamefont {Steinberg}},
  \bibinfo {author} {\bibfnamefont {P.}~\bibnamefont {Jarillo-Herrero}}, \ and\
  \bibinfo {author} {\bibfnamefont {N.}~\bibnamefont {Gedik}},\ }\href
  {\doibase 10.1126/science.1239834} {\bibfield  {journal} {\bibinfo  {journal}
  {Science}\ }\textbf {\bibinfo {volume} {342}},\ \bibinfo {pages} {453}
  (\bibinfo {year} {2013})}\BibitemShut {NoStop}%
\bibitem [{\citenamefont {Kitagawa}\ \emph {et~al.}(2011)\citenamefont
  {Kitagawa}, \citenamefont {Oka}, \citenamefont {Brataas}, \citenamefont
  {Fu},\ and\ \citenamefont {Demler}}]{Kitagawa2011}%
  \BibitemOpen
  \bibfield  {author} {\bibinfo {author} {\bibfnamefont {T.}~\bibnamefont
  {Kitagawa}}, \bibinfo {author} {\bibfnamefont {T.}~\bibnamefont {Oka}},
  \bibinfo {author} {\bibfnamefont {A.}~\bibnamefont {Brataas}}, \bibinfo
  {author} {\bibfnamefont {L.}~\bibnamefont {Fu}}, \ and\ \bibinfo {author}
  {\bibfnamefont {E.}~\bibnamefont {Demler}},\ }\href {\doibase
  10.1103/PhysRevB.84.235108} {\bibfield  {journal} {\bibinfo  {journal}
  {Physical Review B}\ }\textbf {\bibinfo {volume} {84}},\ \bibinfo {pages}
  {235108} (\bibinfo {year} {2011})}\BibitemShut {NoStop}%
\bibitem [{\citenamefont {Sato}\ and\ \citenamefont {Ando}(2017)}]{Sato_2017}%
  \BibitemOpen
  \bibfield  {author} {\bibinfo {author} {\bibfnamefont {M.}~\bibnamefont
  {Sato}}\ and\ \bibinfo {author} {\bibfnamefont {Y.}~\bibnamefont {Ando}},\
  }\href {\doibase https://doi.org/10.1088/1361-6633/aa6ac7} {\bibfield
  {journal} {\bibinfo  {journal} {Reports on Progress in Physics}\ }\textbf
  {\bibinfo {volume} {80}},\ \bibinfo {pages} {076501} (\bibinfo {year}
  {2017})}\BibitemShut {NoStop}%
\bibitem [{\citenamefont {Jiang}\ \emph {et~al.}(2011)\citenamefont {Jiang},
  \citenamefont {Kitagawa}, \citenamefont {Alicea}, \citenamefont {Akhmerov},
  \citenamefont {Pekker}, \citenamefont {Refael}, \citenamefont {Cirac},
  \citenamefont {Demler}, \citenamefont {Lukin},\ and\ \citenamefont
  {Zoller}}]{Jiang_2011}%
  \BibitemOpen
  \bibfield  {author} {\bibinfo {author} {\bibfnamefont {L.}~\bibnamefont
  {Jiang}}, \bibinfo {author} {\bibfnamefont {T.}~\bibnamefont {Kitagawa}},
  \bibinfo {author} {\bibfnamefont {J.}~\bibnamefont {Alicea}}, \bibinfo
  {author} {\bibfnamefont {A.~R.}\ \bibnamefont {Akhmerov}}, \bibinfo {author}
  {\bibfnamefont {D.}~\bibnamefont {Pekker}}, \bibinfo {author} {\bibfnamefont
  {G.}~\bibnamefont {Refael}}, \bibinfo {author} {\bibfnamefont {J.~I.}\
  \bibnamefont {Cirac}}, \bibinfo {author} {\bibfnamefont {E.}~\bibnamefont
  {Demler}}, \bibinfo {author} {\bibfnamefont {M.~D.}\ \bibnamefont {Lukin}}, \
  and\ \bibinfo {author} {\bibfnamefont {P.}~\bibnamefont {Zoller}},\
  }\href@noop {} {\bibfield  {journal} {\bibinfo  {journal} {Physical Review
  Letters}\ }\textbf {\bibinfo {volume} {106}} (\bibinfo {year}
  {2011})}\BibitemShut {NoStop}%
\bibitem [{\citenamefont {Reynoso}\ and\ \citenamefont
  {Frustaglia}(2013)}]{Reynoso2013}%
  \BibitemOpen
  \bibfield  {author} {\bibinfo {author} {\bibfnamefont {A.~A.}\ \bibnamefont
  {Reynoso}}\ and\ \bibinfo {author} {\bibfnamefont {D.}~\bibnamefont
  {Frustaglia}},\ }\href {\doibase https://doi.org/10.1103/PhysRevB.87.115420}
  {\bibfield  {journal} {\bibinfo  {journal} {Physical Review B}\ }\textbf
  {\bibinfo {volume} {87}},\ \bibinfo {pages} {115420} (\bibinfo {year}
  {2013})}\BibitemShut {NoStop}%
\bibitem [{\citenamefont {Thakurathi}\ \emph {et~al.}(2013)\citenamefont
  {Thakurathi}, \citenamefont {Patel}, \citenamefont {Sen},\ and\ \citenamefont
  {Dutta}}]{Thakurathi2013}%
  \BibitemOpen
  \bibfield  {author} {\bibinfo {author} {\bibfnamefont {M.}~\bibnamefont
  {Thakurathi}}, \bibinfo {author} {\bibfnamefont {A.~A.}\ \bibnamefont
  {Patel}}, \bibinfo {author} {\bibfnamefont {D.}~\bibnamefont {Sen}}, \ and\
  \bibinfo {author} {\bibfnamefont {A.}~\bibnamefont {Dutta}},\ }\href
  {\doibase https://doi.org/10.1103/PhysRevB.88.155133} {\bibfield  {journal}
  {\bibinfo  {journal} {Physical Review B}\ }\textbf {\bibinfo {volume} {88}},\
  \bibinfo {pages} {155133} (\bibinfo {year} {2013})}\BibitemShut {NoStop}%
\bibitem [{\citenamefont {Liu}\ \emph {et~al.}(2013{\natexlab{a}})\citenamefont
  {Liu}, \citenamefont {Shan}, \citenamefont {Yao}, \citenamefont {Yao},\ and\
  \citenamefont {Xiao}}]{Liu_2013}%
  \BibitemOpen
  \bibfield  {author} {\bibinfo {author} {\bibfnamefont {G.-B.}\ \bibnamefont
  {Liu}}, \bibinfo {author} {\bibfnamefont {W.-Y.}\ \bibnamefont {Shan}},
  \bibinfo {author} {\bibfnamefont {Y.}~\bibnamefont {Yao}}, \bibinfo {author}
  {\bibfnamefont {W.}~\bibnamefont {Yao}}, \ and\ \bibinfo {author}
  {\bibfnamefont {D.}~\bibnamefont {Xiao}},\ }\href {\doibase
  https://doi.org/10.1103/PhysRevB.88.085433} {\bibfield  {journal} {\bibinfo
  {journal} {Physical Review B}\ }\textbf {\bibinfo {volume} {88}},\ \bibinfo
  {pages} {085433} (\bibinfo {year} {2013}{\natexlab{a}})}\BibitemShut
  {NoStop}%
\bibitem [{\citenamefont {Kundu}\ and\ \citenamefont
  {Seradjeh}(2013)}]{Kundu2013}%
  \BibitemOpen
  \bibfield  {author} {\bibinfo {author} {\bibfnamefont {A.}~\bibnamefont
  {Kundu}}\ and\ \bibinfo {author} {\bibfnamefont {B.}~\bibnamefont
  {Seradjeh}},\ }\href {\doibase 10.1103/PhysRevLett.111.136402} {\bibfield
  {journal} {\bibinfo  {journal} {Physical Review Letters}\ }\textbf {\bibinfo
  {volume} {111}},\ \bibinfo {pages} {136402} (\bibinfo {year}
  {2013})}\BibitemShut {NoStop}%
\bibitem [{\citenamefont {Thakurathi}\ \emph {et~al.}(2017)\citenamefont
  {Thakurathi}, \citenamefont {Loss},\ and\ \citenamefont
  {Klinovaja}}]{Thakurathi_2017}%
  \BibitemOpen
  \bibfield  {author} {\bibinfo {author} {\bibfnamefont {M.}~\bibnamefont
  {Thakurathi}}, \bibinfo {author} {\bibfnamefont {D.}~\bibnamefont {Loss}}, \
  and\ \bibinfo {author} {\bibfnamefont {J.}~\bibnamefont {Klinovaja}},\ }\href
  {\doibase https://doi.org/10.1103/PhysRevB.95.155407} {\bibfield  {journal}
  {\bibinfo  {journal} {Physical Review B}\ }\textbf {\bibinfo {volume} {95}},\
  \bibinfo {pages} {155407} (\bibinfo {year} {2017})}\BibitemShut {NoStop}%
\bibitem [{\citenamefont {Thakurathi}\ \emph {et~al.}(2014)\citenamefont
  {Thakurathi}, \citenamefont {Sengupta},\ and\ \citenamefont
  {Sen}}]{Thakurathi2014}%
  \BibitemOpen
  \bibfield  {author} {\bibinfo {author} {\bibfnamefont {M.}~\bibnamefont
  {Thakurathi}}, \bibinfo {author} {\bibfnamefont {K.}~\bibnamefont
  {Sengupta}}, \ and\ \bibinfo {author} {\bibfnamefont {D.}~\bibnamefont
  {Sen}},\ }\href {\doibase https://doi.org/10.1103/PhysRevB.89.235434}
  {\bibfield  {journal} {\bibinfo  {journal} {Physical Review B}\ }\textbf
  {\bibinfo {volume} {89}},\ \bibinfo {pages} {235434} (\bibinfo {year}
  {2014})}\BibitemShut {NoStop}%
\bibitem [{\citenamefont {Poudel}\ \emph {et~al.}(2015)\citenamefont {Poudel},
  \citenamefont {Ortiz},\ and\ \citenamefont {Viola}}]{Poudel2015}%
  \BibitemOpen
  \bibfield  {author} {\bibinfo {author} {\bibfnamefont {A.}~\bibnamefont
  {Poudel}}, \bibinfo {author} {\bibfnamefont {G.}~\bibnamefont {Ortiz}}, \
  and\ \bibinfo {author} {\bibfnamefont {L.}~\bibnamefont {Viola}},\ }\href
  {\doibase 10.1209/0295-5075/110/17004} {\bibfield  {journal} {\bibinfo
  {journal} {{EPL} (Europhysics Letters)}\ }\textbf {\bibinfo {volume} {110}},\
  \bibinfo {pages} {17004} (\bibinfo {year} {2015})}\BibitemShut {NoStop}%
\bibitem [{\citenamefont {Takasan}\ \emph {et~al.}(2017)\citenamefont
  {Takasan}, \citenamefont {Daido}, \citenamefont {Kawakami},\ and\
  \citenamefont {Yanase}}]{Takasan_2017}%
  \BibitemOpen
  \bibfield  {author} {\bibinfo {author} {\bibfnamefont {K.}~\bibnamefont
  {Takasan}}, \bibinfo {author} {\bibfnamefont {A.}~\bibnamefont {Daido}},
  \bibinfo {author} {\bibfnamefont {N.}~\bibnamefont {Kawakami}}, \ and\
  \bibinfo {author} {\bibfnamefont {Y.}~\bibnamefont {Yanase}},\ }\href
  {\doibase https://doi.org/10.1103/PhysRevB.95.134508} {\bibfield  {journal}
  {\bibinfo  {journal} {Physical Review B}\ }\textbf {\bibinfo {volume} {95}},\
  \bibinfo {pages} {134508} (\bibinfo {year} {2017})}\BibitemShut {NoStop}%
\bibitem [{\citenamefont {Plekhanov}\ \emph {et~al.}(2019)\citenamefont
  {Plekhanov}, \citenamefont {Thakurathi}, \citenamefont {Loss},\ and\
  \citenamefont {Klinovaja}}]{Plekhanov_2019}%
  \BibitemOpen
  \bibfield  {author} {\bibinfo {author} {\bibfnamefont {K.}~\bibnamefont
  {Plekhanov}}, \bibinfo {author} {\bibfnamefont {M.}~\bibnamefont
  {Thakurathi}}, \bibinfo {author} {\bibfnamefont {D.}~\bibnamefont {Loss}}, \
  and\ \bibinfo {author} {\bibfnamefont {J.}~\bibnamefont {Klinovaja}},\ }\href
  {\doibase https://doi.org/10.1103/PhysRevResearch.1.032013} {\bibfield
  {journal} {\bibinfo  {journal} {Physical Review Research}\ }\textbf {\bibinfo
  {volume} {1}},\ \bibinfo {pages} {032013} (\bibinfo {year}
  {2019})}\BibitemShut {NoStop}%
\bibitem [{\citenamefont {Deng}\ \emph {et~al.}(2014)\citenamefont {Deng},
  \citenamefont {Ortiz}, \citenamefont {Poudel},\ and\ \citenamefont
  {Viola}}]{Deng2014}%
  \BibitemOpen
  \bibfield  {author} {\bibinfo {author} {\bibfnamefont {S.}~\bibnamefont
  {Deng}}, \bibinfo {author} {\bibfnamefont {G.}~\bibnamefont {Ortiz}},
  \bibinfo {author} {\bibfnamefont {A.}~\bibnamefont {Poudel}}, \ and\ \bibinfo
  {author} {\bibfnamefont {L.}~\bibnamefont {Viola}},\ }\href {\doibase
  https://doi.org/10.1103/PhysRevB.89.140507} {\bibfield  {journal} {\bibinfo
  {journal} {Physical Review B}\ }\textbf {\bibinfo {volume} {89}},\ \bibinfo
  {pages} {140507} (\bibinfo {year} {2014})}\BibitemShut {NoStop}%
\bibitem [{\citenamefont {Bir}\ and\ \citenamefont
  {Pikus}(1974)}]{BirPikusBook}%
  \BibitemOpen
  \bibfield  {author} {\bibinfo {author} {\bibfnamefont {G.~L.}\ \bibnamefont
  {Bir}}\ and\ \bibinfo {author} {\bibfnamefont {G.~E.}\ \bibnamefont
  {Pikus}},\ }\href@noop {} {\emph {\bibinfo {title} {{Symmetry and
  Strain-induced Effects in Semiconductors}}}}\ (\bibinfo  {publisher}
  {Wiley},\ \bibinfo {address} {New York},\ \bibinfo {year} {1974})\BibitemShut
  {NoStop}%
\bibitem [{\citenamefont {Winkler}(2003)}]{winklerbook}%
  \BibitemOpen
  \bibfield  {author} {\bibinfo {author} {\bibfnamefont {R.}~\bibnamefont
  {Winkler}},\ }\href {\doibase 10.1007/b13586} {\emph {\bibinfo {title}
  {Spin--Orbit Coupling Effects in Two-Dimensional Electron and Hole
  Systems}}},\ \bibinfo {series} {Springer Tracts in Modern Physics}, Vol.\
  \bibinfo {volume} {191}\ (\bibinfo  {publisher} {Springer Berlin
  Heidelberg},\ \bibinfo {address} {Berlin, Heidelberg},\ \bibinfo {year}
  {2003})\ p.\ \bibinfo {pages} {228}\BibitemShut {NoStop}%
\bibitem [{\citenamefont {Bychkov}\ and\ \citenamefont
  {Rashba}(1984)}]{rashba_1}%
  \BibitemOpen
  \bibfield  {author} {\bibinfo {author} {\bibfnamefont {Y.}~\bibnamefont
  {Bychkov}}\ and\ \bibinfo {author} {\bibfnamefont {E.~I.}\ \bibnamefont
  {Rashba}},\ }\href@noop {} {\bibfield  {journal} {\bibinfo  {journal} {JETP
  Lett.}\ }\textbf {\bibinfo {volume} {39}},\ \bibinfo {pages} {78} (\bibinfo
  {year} {1984})}\BibitemShut {NoStop}%
\bibitem [{\citenamefont {Rashba}\ and\ \citenamefont
  {Sheka}(2015)}]{Rashba2015SymmetryOE}%
  \BibitemOpen
  \bibfield  {author} {\bibinfo {author} {\bibfnamefont {E.}~\bibnamefont
  {Rashba}}\ and\ \bibinfo {author} {\bibnamefont {Sheka}},\ }\href@noop {}
  {\enquote {\bibinfo {title} {{Symmetry of Energy Bands in Crystals of
  Wurtzite Type II. Symmetry of Bands with Spin-Orbit Interaction Included}},}\
  }\bibinfo {howpublished}
  {\url{http://nrs.harvard.edu/urn-3:HUL.InstRepos:29426010}} (\bibinfo {year}
  {2015})\BibitemShut {NoStop}%
\bibitem [{\citenamefont {Dresselhaus}(1955)}]{Dresselhaus1955}%
  \BibitemOpen
  \bibfield  {author} {\bibinfo {author} {\bibfnamefont {G.}~\bibnamefont
  {Dresselhaus}},\ }\href {\doibase 10.1103/PhysRev.100.580} {\bibfield
  {journal} {\bibinfo  {journal} {Physical Review}\ }\textbf {\bibinfo {volume}
  {100}},\ \bibinfo {pages} {580} (\bibinfo {year} {1955})}\BibitemShut
  {NoStop}%
\bibitem [{Note1()}]{Note1}%
  \BibitemOpen
  \bibinfo {note} {Tensor convention: $\tau _z\otimes \sigma _i = \begin
  {pmatrix} \sigma _i & \vb {0}_2\\ \vb {0}_2 & -\sigma _i \end
  {pmatrix}$}\BibitemShut {NoStop}%
\bibitem [{\citenamefont {Sheremet}\ \emph {et~al.}(2016)\citenamefont
  {Sheremet}, \citenamefont {Kibis}, \citenamefont {Kavokin},\ and\
  \citenamefont {Shelykh}}]{Sheremet2016}%
  \BibitemOpen
  \bibfield  {author} {\bibinfo {author} {\bibfnamefont {A.~S.}\ \bibnamefont
  {Sheremet}}, \bibinfo {author} {\bibfnamefont {O.~V.}\ \bibnamefont {Kibis}},
  \bibinfo {author} {\bibfnamefont {A.~V.}\ \bibnamefont {Kavokin}}, \ and\
  \bibinfo {author} {\bibfnamefont {I.~A.}\ \bibnamefont {Shelykh}},\ }\href
  {\doibase 10.1103/PhysRevB.93.165307} {\bibfield  {journal} {\bibinfo
  {journal} {Physical Review B}\ }\textbf {\bibinfo {volume} {93}},\ \bibinfo
  {pages} {165307} (\bibinfo {year} {2016})}\BibitemShut {NoStop}%
\bibitem [{\citenamefont {Liu}\ \emph {et~al.}(2013{\natexlab{b}})\citenamefont
  {Liu}, \citenamefont {Levchenko},\ and\ \citenamefont
  {Baranger}}]{Liu_2013_Floq}%
  \BibitemOpen
  \bibfield  {author} {\bibinfo {author} {\bibfnamefont {D.~E.}\ \bibnamefont
  {Liu}}, \bibinfo {author} {\bibfnamefont {A.}~\bibnamefont {Levchenko}}, \
  and\ \bibinfo {author} {\bibfnamefont {H.~U.}\ \bibnamefont {Baranger}},\
  }\href {\doibase https://doi.org/10.1103/PhysRevLett.111.047002} {\bibfield
  {journal} {\bibinfo  {journal} {Physical Review Letters}\ }\textbf {\bibinfo
  {volume} {111}},\ \bibinfo {pages} {047002} (\bibinfo {year}
  {2013}{\natexlab{b}})}\BibitemShut {NoStop}%
\bibitem [{\citenamefont {Alicea}(2012)}]{Alicea_2012}%
  \BibitemOpen
  \bibfield  {author} {\bibinfo {author} {\bibfnamefont {J.}~\bibnamefont
  {Alicea}},\ }\href {\doibase https://doi.org/10.1088/0034-4885/75/7/076501}
  {\bibfield  {journal} {\bibinfo  {journal} {Reports on Progress in Physics}\
  }\textbf {\bibinfo {volume} {75}},\ \bibinfo {pages} {076501} (\bibinfo
  {year} {2012})}\BibitemShut {NoStop}%
\bibitem [{\citenamefont {Sau}\ \emph {et~al.}(2010)\citenamefont {Sau},
  \citenamefont {Lutchyn}, \citenamefont {Tewari},\ and\ \citenamefont
  {Sarma}}]{Sau_2010}%
  \BibitemOpen
  \bibfield  {author} {\bibinfo {author} {\bibfnamefont {J.~D.}\ \bibnamefont
  {Sau}}, \bibinfo {author} {\bibfnamefont {R.~M.}\ \bibnamefont {Lutchyn}},
  \bibinfo {author} {\bibfnamefont {S.}~\bibnamefont {Tewari}}, \ and\ \bibinfo
  {author} {\bibfnamefont {S.~D.}\ \bibnamefont {Sarma}},\ }\href {\doibase
  https://doi.org/10.1103/PhysRevLett.104.040502} {\bibfield  {journal}
  {\bibinfo  {journal} {Physical Review Letters}\ }\textbf {\bibinfo {volume}
  {104}},\ \bibinfo {pages} {040502} (\bibinfo {year} {2010})}\BibitemShut
  {NoStop}%
\bibitem [{\citenamefont {Thouless}\ \emph {et~al.}(1982)\citenamefont
  {Thouless}, \citenamefont {Kohmoto}, \citenamefont {Nightingale},\ and\
  \citenamefont {{Den Nijs}}}]{Thouless1982a}%
  \BibitemOpen
  \bibfield  {author} {\bibinfo {author} {\bibfnamefont {D.~J.}\ \bibnamefont
  {Thouless}}, \bibinfo {author} {\bibfnamefont {M.}~\bibnamefont {Kohmoto}},
  \bibinfo {author} {\bibfnamefont {M.~P.}\ \bibnamefont {Nightingale}}, \ and\
  \bibinfo {author} {\bibfnamefont {M.}~\bibnamefont {{Den Nijs}}},\ }\href
  {\doibase 10.1103/PhysRevLett.49.405} {\bibfield  {journal} {\bibinfo
  {journal} {Physical Review Letters}\ }\textbf {\bibinfo {volume} {49}},\
  \bibinfo {pages} {405} (\bibinfo {year} {1982})}\BibitemShut {NoStop}%
\bibitem [{\citenamefont {Berry}(1984)}]{Berry1984}%
  \BibitemOpen
  \bibfield  {author} {\bibinfo {author} {\bibfnamefont {M.~V.}\ \bibnamefont
  {Berry}},\ }\href {\doibase 10.1098/rspa.1984.0023} {\bibfield  {journal}
  {\bibinfo  {journal} {Proceedings of the Royal Society A}\ }\textbf {\bibinfo
  {volume} {392}},\ \bibinfo {pages} {45} (\bibinfo {year} {1984})}\BibitemShut
  {NoStop}%
\bibitem [{\citenamefont {Simon}(1983)}]{Simon1983}%
  \BibitemOpen
  \bibfield  {author} {\bibinfo {author} {\bibfnamefont {B.}~\bibnamefont
  {Simon}},\ }\href {\doibase 10.1103/physrevlett.51.2167} {\bibfield
  {journal} {\bibinfo  {journal} {Physical Review Letters}\ }\textbf {\bibinfo
  {volume} {51}},\ \bibinfo {pages} {2167} (\bibinfo {year}
  {1983})}\BibitemShut {NoStop}%
\bibitem [{\citenamefont {Schliemann}\ \emph {et~al.}(2003)\citenamefont
  {Schliemann}, \citenamefont {Egues},\ and\ \citenamefont
  {Loss}}]{Schliemann2003}%
  \BibitemOpen
  \bibfield  {author} {\bibinfo {author} {\bibfnamefont {J.}~\bibnamefont
  {Schliemann}}, \bibinfo {author} {\bibfnamefont {J.~C.}\ \bibnamefont
  {Egues}}, \ and\ \bibinfo {author} {\bibfnamefont {D.}~\bibnamefont {Loss}},\
  }\href {\doibase 10.1103/PhysRevLett.90.146801} {\bibfield  {journal}
  {\bibinfo  {journal} {Physical Review Letters}\ }\textbf {\bibinfo {volume}
  {90}},\ \bibinfo {pages} {146801} (\bibinfo {year} {2003})}\BibitemShut
  {NoStop}%
\bibitem [{\citenamefont {Koralek}\ \emph {et~al.}(2009)\citenamefont
  {Koralek}, \citenamefont {Weber}, \citenamefont {Orenstein}, \citenamefont
  {Bernevig}, \citenamefont {Zhang}, \citenamefont {Mack},\ and\ \citenamefont
  {Awschalom}}]{Koralek2009}%
  \BibitemOpen
  \bibfield  {author} {\bibinfo {author} {\bibfnamefont {J.~D.}\ \bibnamefont
  {Koralek}}, \bibinfo {author} {\bibfnamefont {C.~P.}\ \bibnamefont {Weber}},
  \bibinfo {author} {\bibfnamefont {J.}~\bibnamefont {Orenstein}}, \bibinfo
  {author} {\bibfnamefont {B.~A.}\ \bibnamefont {Bernevig}}, \bibinfo {author}
  {\bibfnamefont {S.-C.}\ \bibnamefont {Zhang}}, \bibinfo {author}
  {\bibfnamefont {S.}~\bibnamefont {Mack}}, \ and\ \bibinfo {author}
  {\bibfnamefont {D.~D.}\ \bibnamefont {Awschalom}},\ }\href
  {http://dx.doi.org/10.1038/nature07871} {\bibfield  {journal} {\bibinfo
  {journal} {Nature}\ }\textbf {\bibinfo {volume} {458}},\ \bibinfo {pages}
  {610} (\bibinfo {year} {2009})}\BibitemShut {NoStop}%
\bibitem [{\citenamefont {Kammermeier}\ \emph {et~al.}(2016)\citenamefont
  {Kammermeier}, \citenamefont {Wenk},\ and\ \citenamefont
  {Schliemann}}]{Kammermeier2016a}%
  \BibitemOpen
  \bibfield  {author} {\bibinfo {author} {\bibfnamefont {M.}~\bibnamefont
  {Kammermeier}}, \bibinfo {author} {\bibfnamefont {P.}~\bibnamefont {Wenk}}, \
  and\ \bibinfo {author} {\bibfnamefont {J.}~\bibnamefont {Schliemann}},\
  }\href {\doibase 10.1103/PhysRevLett.117.236801} {\bibfield  {journal}
  {\bibinfo  {journal} {Physical Review Letters}\ }\textbf {\bibinfo {volume}
  {117}},\ \bibinfo {pages} {236801} (\bibinfo {year} {2016})}\BibitemShut
  {NoStop}%
\bibitem [{\citenamefont {Schliemann}(2017)}]{Schliemann_2017}%
  \BibitemOpen
  \bibfield  {author} {\bibinfo {author} {\bibfnamefont {J.}~\bibnamefont
  {Schliemann}},\ }\href {\doibase 10.1103/revmodphys.89.011001} {\bibfield
  {journal} {\bibinfo  {journal} {Reviews of Modern Physics}\ }\textbf
  {\bibinfo {volume} {89}},\ \bibinfo {pages} {011001} (\bibinfo {year}
  {2017})}\BibitemShut {NoStop}%
\bibitem [{\citenamefont {Oshima}\ \emph {et~al.}(2022)\citenamefont {Oshima},
  \citenamefont {Ikegaya}, \citenamefont {Schnyder},\ and\ \citenamefont
  {Tanaka}}]{Oshima_2022}%
  \BibitemOpen
  \bibfield  {author} {\bibinfo {author} {\bibfnamefont {D.}~\bibnamefont
  {Oshima}}, \bibinfo {author} {\bibfnamefont {S.}~\bibnamefont {Ikegaya}},
  \bibinfo {author} {\bibfnamefont {A.~P.}\ \bibnamefont {Schnyder}}, \ and\
  \bibinfo {author} {\bibfnamefont {Y.}~\bibnamefont {Tanaka}},\ }\href
  {\doibase 10.1103/PhysRevResearch.4.L022051} {\bibfield  {journal} {\bibinfo
  {journal} {Phys. Rev. Research}\ }\textbf {\bibinfo {volume} {4}},\ \bibinfo
  {pages} {L022051} (\bibinfo {year} {2022})}\BibitemShut {NoStop}%
\bibitem [{\citenamefont {Altland}\ and\ \citenamefont
  {Zirnbauer}(1997)}]{Altland_1997}%
  \BibitemOpen
  \bibfield  {author} {\bibinfo {author} {\bibfnamefont {A.}~\bibnamefont
  {Altland}}\ and\ \bibinfo {author} {\bibfnamefont {M.~R.}\ \bibnamefont
  {Zirnbauer}},\ }\href {\doibase 10.1103/PhysRevB.55.1142} {\bibfield
  {journal} {\bibinfo  {journal} {Physical Review B}\ }\textbf {\bibinfo
  {volume} {55}},\ \bibinfo {pages} {1142} (\bibinfo {year}
  {1997})}\BibitemShut {NoStop}%
\bibitem [{\citenamefont {Mishmash}\ \emph {et~al.}(2016)\citenamefont
  {Mishmash}, \citenamefont {Aasen}, \citenamefont {Higginbotham},\ and\
  \citenamefont {Alicea}}]{Mishmash2016}%
  \BibitemOpen
  \bibfield  {author} {\bibinfo {author} {\bibfnamefont {R.~V.}\ \bibnamefont
  {Mishmash}}, \bibinfo {author} {\bibfnamefont {D.}~\bibnamefont {Aasen}},
  \bibinfo {author} {\bibfnamefont {A.~P.}\ \bibnamefont {Higginbotham}}, \
  and\ \bibinfo {author} {\bibfnamefont {J.}~\bibnamefont {Alicea}},\ }\href
  {\doibase 10.1103/physrevb.93.245404} {\bibfield  {journal} {\bibinfo
  {journal} {Physical Review B}\ }\textbf {\bibinfo {volume} {93}},\ \bibinfo
  {pages} {245404} (\bibinfo {year} {2016})}\BibitemShut {NoStop}%
\bibitem [{\citenamefont {Leumer}\ \emph {et~al.}(2021)\citenamefont {Leumer},
  \citenamefont {Grifoni}, \citenamefont {Muralidharan},\ and\ \citenamefont
  {Marganska}}]{Leumer_2021}%
  \BibitemOpen
  \bibfield  {author} {\bibinfo {author} {\bibfnamefont {N.}~\bibnamefont
  {Leumer}}, \bibinfo {author} {\bibfnamefont {M.}~\bibnamefont {Grifoni}},
  \bibinfo {author} {\bibfnamefont {B.}~\bibnamefont {Muralidharan}}, \ and\
  \bibinfo {author} {\bibfnamefont {M.}~\bibnamefont {Marganska}},\ }\href
  {\doibase https://doi.org/10.1103/PhysRevB.103.165432} {\bibfield  {journal}
  {\bibinfo  {journal} {Physical Review B}\ }\textbf {\bibinfo {volume}
  {103}},\ \bibinfo {pages} {165432} (\bibinfo {year} {2021})}\BibitemShut
  {NoStop}%
\bibitem [{\citenamefont {Qu}\ \emph {et~al.}(2016)\citenamefont {Qu},
  \citenamefont {van Veen}, \citenamefont {de~Vries}, \citenamefont {Beukman},
  \citenamefont {Wimmer}, \citenamefont {Yi}, \citenamefont {Kiselev},
  \citenamefont {Nguyen}, \citenamefont {Sokolich}, \citenamefont {Manfra},
  \citenamefont {Nichele}, \citenamefont {Marcus},\ and\ \citenamefont
  {Kouwenhoven}}]{Qu_2016}%
  \BibitemOpen
  \bibfield  {author} {\bibinfo {author} {\bibfnamefont {F.}~\bibnamefont
  {Qu}}, \bibinfo {author} {\bibfnamefont {J.}~\bibnamefont {van Veen}},
  \bibinfo {author} {\bibfnamefont {F.~K.}\ \bibnamefont {de~Vries}}, \bibinfo
  {author} {\bibfnamefont {A.~J.~A.}\ \bibnamefont {Beukman}}, \bibinfo
  {author} {\bibfnamefont {M.}~\bibnamefont {Wimmer}}, \bibinfo {author}
  {\bibfnamefont {W.}~\bibnamefont {Yi}}, \bibinfo {author} {\bibfnamefont
  {A.~A.}\ \bibnamefont {Kiselev}}, \bibinfo {author} {\bibfnamefont {B.-M.}\
  \bibnamefont {Nguyen}}, \bibinfo {author} {\bibfnamefont {M.}~\bibnamefont
  {Sokolich}}, \bibinfo {author} {\bibfnamefont {M.~J.}\ \bibnamefont
  {Manfra}}, \bibinfo {author} {\bibfnamefont {F.}~\bibnamefont {Nichele}},
  \bibinfo {author} {\bibfnamefont {C.~M.}\ \bibnamefont {Marcus}}, \ and\
  \bibinfo {author} {\bibfnamefont {L.~P.}\ \bibnamefont {Kouwenhoven}},\
  }\href {\doibase https://doi.org/10.1021/acs.nanolett.6b03297} {\bibfield
  {journal} {\bibinfo  {journal} {Nano Letters}\ }\textbf {\bibinfo {volume}
  {16}},\ \bibinfo {pages} {6509} (\bibinfo {year} {2016})}\BibitemShut
  {NoStop}%
\bibitem [{\citenamefont {Xi}\ \emph {et~al.}(2015)\citenamefont {Xi},
  \citenamefont {Wang}, \citenamefont {Zhao}, \citenamefont {Park},
  \citenamefont {Law}, \citenamefont {Berger}, \citenamefont {Forr{\'{o}}},
  \citenamefont {Shan},\ and\ \citenamefont {Mak}}]{Xi_2016}%
  \BibitemOpen
  \bibfield  {author} {\bibinfo {author} {\bibfnamefont {X.}~\bibnamefont
  {Xi}}, \bibinfo {author} {\bibfnamefont {Z.}~\bibnamefont {Wang}}, \bibinfo
  {author} {\bibfnamefont {W.}~\bibnamefont {Zhao}}, \bibinfo {author}
  {\bibfnamefont {J.-H.}\ \bibnamefont {Park}}, \bibinfo {author}
  {\bibfnamefont {K.~T.}\ \bibnamefont {Law}}, \bibinfo {author} {\bibfnamefont
  {H.}~\bibnamefont {Berger}}, \bibinfo {author} {\bibfnamefont
  {L.}~\bibnamefont {Forr{\'{o}}}}, \bibinfo {author} {\bibfnamefont
  {J.}~\bibnamefont {Shan}}, \ and\ \bibinfo {author} {\bibfnamefont {K.~F.}\
  \bibnamefont {Mak}},\ }\href {\doibase 10.1038/nphys3538} {\bibfield
  {journal} {\bibinfo  {journal} {Nature Physics}\ }\textbf {\bibinfo {volume}
  {12}},\ \bibinfo {pages} {139} (\bibinfo {year} {2015})}\BibitemShut
  {NoStop}%
\bibitem [{\citenamefont {Kuzmanović}\ \emph {et~al.}(2021)\citenamefont
  {Kuzmanović}, \citenamefont {Dvir}, \citenamefont {LeBoeuf}, \citenamefont
  {Ilić}, \citenamefont {Möckli}, \citenamefont {Haim}, \citenamefont
  {Kraemer}, \citenamefont {Khodas}, \citenamefont {Houzet}, \citenamefont
  {Meyer}, \citenamefont {Aprili}, \citenamefont {Steinberg},\ and\
  \citenamefont {Quay}}]{Kuzmanovic_2021}%
  \BibitemOpen
  \bibfield  {author} {\bibinfo {author} {\bibfnamefont {M.}~\bibnamefont
  {Kuzmanović}}, \bibinfo {author} {\bibfnamefont {T.}~\bibnamefont {Dvir}},
  \bibinfo {author} {\bibfnamefont {D.}~\bibnamefont {LeBoeuf}}, \bibinfo
  {author} {\bibfnamefont {S.}~\bibnamefont {Ilić}}, \bibinfo {author}
  {\bibfnamefont {D.}~\bibnamefont {Möckli}}, \bibinfo {author} {\bibfnamefont
  {M.}~\bibnamefont {Haim}}, \bibinfo {author} {\bibfnamefont {S.}~\bibnamefont
  {Kraemer}}, \bibinfo {author} {\bibfnamefont {M.}~\bibnamefont {Khodas}},
  \bibinfo {author} {\bibfnamefont {M.}~\bibnamefont {Houzet}}, \bibinfo
  {author} {\bibfnamefont {J.~S.}\ \bibnamefont {Meyer}}, \bibinfo {author}
  {\bibfnamefont {M.}~\bibnamefont {Aprili}}, \bibinfo {author} {\bibfnamefont
  {H.}~\bibnamefont {Steinberg}}, \ and\ \bibinfo {author} {\bibfnamefont
  {C.~H.~L.}\ \bibnamefont {Quay}},\ }\href@noop {} {\enquote {\bibinfo {title}
  {{Tunneling spectroscopy of few-monolayer NbSe$_2$ in high magnetic field:
  Ising protection and triplet superconductivity}},}\ } (\bibinfo {year}
  {2021}),\ \Eprint {http://arxiv.org/abs/2104.00328} {arXiv:2104.00328
  [cond-mat.supr-con]} \BibitemShut {NoStop}%
\bibitem [{\citenamefont {Mikami}\ \emph {et~al.}(2016)\citenamefont {Mikami},
  \citenamefont {Kitamura}, \citenamefont {Yasuda}, \citenamefont {Tsuji},
  \citenamefont {Oka},\ and\ \citenamefont {Aoki}}]{Mikami2016}%
  \BibitemOpen
  \bibfield  {author} {\bibinfo {author} {\bibfnamefont {T.}~\bibnamefont
  {Mikami}}, \bibinfo {author} {\bibfnamefont {S.}~\bibnamefont {Kitamura}},
  \bibinfo {author} {\bibfnamefont {K.}~\bibnamefont {Yasuda}}, \bibinfo
  {author} {\bibfnamefont {N.}~\bibnamefont {Tsuji}}, \bibinfo {author}
  {\bibfnamefont {T.}~\bibnamefont {Oka}}, \ and\ \bibinfo {author}
  {\bibfnamefont {H.}~\bibnamefont {Aoki}},\ }\href {\doibase
  10.1103/physrevb.93.144307} {\bibfield  {journal} {\bibinfo  {journal}
  {Physical Review B}\ }\textbf {\bibinfo {volume} {93}},\ \bibinfo {pages}
  {144307} (\bibinfo {year} {2016})}\BibitemShut {NoStop}%
\bibitem [{\citenamefont {Floquet}(1883)}]{Floquet_1883}%
  \BibitemOpen
  \bibfield  {author} {\bibinfo {author} {\bibfnamefont {G.}~\bibnamefont
  {Floquet}},\ }\href {\doibase 10.24033/asens.220} {\bibfield  {journal}
  {\bibinfo  {journal} {Annales scientifiques de l{\'{E}}cole normale
  sup{\'{e}}rieure}\ }\textbf {\bibinfo {volume} {12}},\ \bibinfo {pages} {47}
  (\bibinfo {year} {1883})}\BibitemShut {NoStop}%
\bibitem [{\citenamefont {Marganska}\ \emph {et~al.}(2018)\citenamefont
  {Marganska}, \citenamefont {Milz}, \citenamefont {Izumida}, \citenamefont
  {Strunk},\ and\ \citenamefont {Grifoni}}]{Marganska_2018}%
  \BibitemOpen
  \bibfield  {author} {\bibinfo {author} {\bibfnamefont {M.}~\bibnamefont
  {Marganska}}, \bibinfo {author} {\bibfnamefont {L.}~\bibnamefont {Milz}},
  \bibinfo {author} {\bibfnamefont {W.}~\bibnamefont {Izumida}}, \bibinfo
  {author} {\bibfnamefont {C.}~\bibnamefont {Strunk}}, \ and\ \bibinfo {author}
  {\bibfnamefont {M.}~\bibnamefont {Grifoni}},\ }\href {\doibase
  https://doi.org/10.1103/PhysRevB.97.075141} {\bibfield  {journal} {\bibinfo
  {journal} {Physical Review B}\ }\textbf {\bibinfo {volume} {97}},\ \bibinfo
  {pages} {075141} (\bibinfo {year} {2018})}\BibitemShut {NoStop}%
\bibitem [{\citenamefont {Deng}\ \emph {et~al.}(2013)\citenamefont {Deng},
  \citenamefont {Ortiz},\ and\ \citenamefont {Viola}}]{Deng_2013}%
  \BibitemOpen
  \bibfield  {author} {\bibinfo {author} {\bibfnamefont {S.}~\bibnamefont
  {Deng}}, \bibinfo {author} {\bibfnamefont {G.}~\bibnamefont {Ortiz}}, \ and\
  \bibinfo {author} {\bibfnamefont {L.}~\bibnamefont {Viola}},\ }\href
  {\doibase http://dx.doi.org/10.1103/PhysRevB.87.205414} {\bibfield  {journal}
  {\bibinfo  {journal} {Physical Review B}\ }\textbf {\bibinfo {volume} {87}},\
  \bibinfo {pages} {205414} (\bibinfo {year} {2013})}\BibitemShut {NoStop}%
\end{thebibliography}%
\end{document}